\documentclass[12pt]{article}

\usepackage{graphicx}
\usepackage[dvips]{color}
\usepackage[sort&compress,super,comma]{natbib}

\makeatletter 
\renewcommand\@biblabel[1]{#1.}
\makeatother

\title{ Dispersion Terms and Analysis of Size- and Charge- 
        Dependence in an Enhanced Poisson-Boltzmann Approach }

\author{ Parimal Kar $^{\dagger}$,
         Max Seel $^{\dagger}$, 
         Ulrich H. E. Hansmann $^{\dagger, \ddagger}$,\\       
         Siegfried H\"ofinger 
         $^{\dagger,}$\footnote{ \scriptsize
                                 Corresponding author.
                                 E-Mail: shoefing@mtu.edu.    
                                 Phone: (906) 487-1496. 
                                 Fax:   (906) 487-2933.
                                 \newline
                                 \hspace*{0.5cm}$^{\dagger}$Department 
                                                       of Physics, 
                                                       Michigan 
                                                       Technological 
                                                       University
                                 \newline
                                 \hspace*{0.5cm}$^{\ddagger}$John von 
                                                       Neumann Institute 
                                                       for Computing, 
                                                       FZ J\"ulich      
                               } \\
       }
\date{ }

\begin{document}
\maketitle
\noindent
{\it Department of Physics, Michigan Technological
     University, 1400 Townsend Drive, Houghton, MI,
     49931-1295, USA, 
     and
     John von Neumann Institute for Computing, Forschungszentrum 
     J\"ulich, 52425 J\"ulich, Germany  }
         
{\scriptsize RECEIVED DATE (to be automatically inserted) }   

\newpage

\begin{abstract}
\noindent
We implement a well-established concept to consider dispersion effects 
within a Poisson-Boltzmann approach of continuum solvation of proteins.
The theoretical framework is particularly suited for boundary element methods. 
Free parameters are determined by comparison to experimental data as well as 
high level Quantum Mechanical reference calculations. The method is general 
and can be easily extended in several directions. The model is tested on 
various chemical substances and found to yield good quality estimates of the 
solvation free energy without obvious indication of any introduced bias. 
Once optimized, the model is applied to a series of proteins and factors 
such as protein size or partial charge assignments are studied.
\end{abstract}

\pagebreak

\section{ Introduction }
The stabilizing effect of water on biomolecules is an intensively studied 
area in contemporary biophysical research. This is because many of the key 
principles governing biological functionality result from the action of the 
solvent, and thus water is often regarded as the ``matrix of life''.

In theoretical work, the important factor ``solvent'' needs to be taken 
into account too, or the studied system will be unphysical. There are two 
main ways of solvent treatment in biophysical research. One is to embed the 
biomolecule of interest into a box of explicit solvent molecules resolved 
into full atomic detail \cite{jorgensen,berendsen,ren}. The alternative form 
is to consider the solvent as structureless continuum and describe the 
response of the environment with implicit solvation methods 
\cite{tomasi,rivail,roux,ooi,cramer}. Much effort has been devoted to 
describing the electrostatic component within implicit solvation models.  
Efficient solutions have become popular in the form of {\it Generalized Born} 
(GB) models \cite{still,case,feig,schaefer} as well as {\it Poisson-Boltzmann} 
models (PB) 
\cite{warwicker,honig,gilson,mccammon,karplus,van_gunsteren,zauhar,juffer}.
Solutions to the PB are computed either by the finite difference
method (FDPB) \cite{warwicker,honig,gilson,mccammon} or by the boundary
element method (PB/BEM) \cite{zauhar,juffer}. Considerable computational
savings are expected from the latter because the problem can be reduced 
from having to solve a volume integral in FDPB to solving a surface 
integral in PB/BEM. Either approach is sensitive to the degree of 
discretization into grid elements or boundary elements \cite{mohan,parimal}.

Aside from the electrostatic component there are also apolar contributions 
to consider \cite{tomasi,zacharias}. Especially in the context of nonpolar 
molecules, such factors often become the dominant terms in the solvation free 
energy. A common way to treat these nonpolar contributions is to introduce a 
SASA-term, which means measuring the solvent accessible surface area (SASA) 
and weighing it with an empirically determined factor. Although commonly 
employed, this procedure has become the subject of intensive debates 
\cite{van_gunsteren2,simonson,su,wagoner}. Not only were SASA terms found to 
be inappropriate for representing the cavitation term \cite{hoefi,wagoner}, 
but also is the weighing factor --- usually associated with surface tension --- 
completely ill-defined in an atomic scale context \cite{mahajan}. While
the short range character of dispersion and repulsion forces occurring at the 
boundary between solute and solvent would imply that SASA can describe these 
kinds of interactions, a recent careful analysis has shown that at least for
dispersion such a relationship is not justified \cite{van_gunsteren2}. 

The discrepancy arising with SASA-terms has been recognized by many groups
and its persistent employment may be largely due to sizeable  cancellation 
of error effects. 
\textcolor{red}{Wagoner and Baker \cite{wagoner} have devided the non-polar 
                contributions into repulsive and attractive components and 
                compared their approach to mean forces obtained from 
                simulation data on explicitly solvated systems. The specific
                role of the volume to account for repulsive interactions 
                (cavitation) was clearly identified. Further inclusion
                of a dispersion term resulted in a satisfactory model of  
                high predictive quality. Levy and Gallicchio devised a 
                similar decomposition into a SASA-dependent cavitation term 
                and a dispersion term within a GB scheme 
                \cite{galli,galli1,levy}. Their model makes use of atomic 
                surface tensions and a rigorous definition of the molecular 
                geometry within the GB framework. Particularly attractive is 
                their efficient implementation and  straightforward 
                interfacing with Molecular Dynamics codes.
                Zacharias has already noted that a decompositon into 
                a dispersion term and a SASA-based cavity term greatly 
                benefits the quality of predictions of apolar solvation 
                \cite{zacharias}. His approach uses distinct surface layers
                for either contribution. Hydration free energies of a series 
                of tested alkanes agreed very well with data from explicit 
                simulations \cite{galli2} and from experiment. The striking
                feature in this approach is the improvement in hydration free 
                energies of cyclic alkanes.
               }
Methodic advancement has recently been reported within the
newest release of AMBER \cite{amber} where GB was augmented by a volume term
\cite{onuvriev} and the inclusion of dispersion terms was found to
significantly improve the general predictive quality of PB. Of particular 
interest are systematic and physics-based decompositions that allow for 
separate consideration of each of the terms involved. In Quantum Mechanics 
(QM) such a technique has long been established with the Polarizable Continuum 
Model (PCM) \cite{tomasi}. It therefore seems advisable to use techniques like 
PCM as a reference system whenever additional method development is performed, 
especially when regarding the multitude of technical dependencies continuum 
solvation models are faced with \cite{mohan,parimal}.

In the present work we describe a systematic process to introduce dispersion
terms in the context of the PB/BEM approach. The PCM model, that treats   
dispersion and repulsion terms from first-principles, is used as a reference 
system along with experimental data. Different ways of calculating 
dispersion-, repulsion contributions in PCM have recently been compared 
\cite{curutchet}. For our purposes the Caillet-Claverie method 
\cite{caillet,claverie} was implemented since it seems to offer a good 
compromise between accuracy and computational overhead. This method was also 
chosen in earlier versions of PCM  \cite{floris} and thus represents a proven 
concept within the BEM framework. The fundamental role of dispersion and the 
potential danger of misinterpreting hydrophobicity related phenomena by   
ignoring it has been underlined recently \cite{choudhury,choudhury1}. 
\textcolor{red}{Given the fundmental nature of hydrophobicity and the
                potential role of dispersion within it together with the 
                current diversity seen in all the explanatory model 
                concepts \cite{choudhury2,zhou} it seems to be necessary to 
                advance all technical refinements to all solvation models 
                (implicit as well as explicit) just to facilitate an eventual 
                understanding of the factors governing these basic
                structure-forming principles.
               }

After determination of appropriate dispersion constants used in the 
Caillet-Claverie approach we apply our model to a series of proteins of 
increasing size. In this way we can analyze the relative contribution of the 
individual terms as a function of system size. Moreover, we have carried out 
semi-empirical calculations on the same series of proteins and can therefore 
compare effects resulting from different charge assignments to each other.
The semi-empirical program LocalSCF \cite{anikin} also allowed for estimation 
of the polarization free energies according to the COSMO model \cite{klamt}, 
which could be readily used for direct comparison to PB/BEM data.

\section{ Methods }
\subsection{ Theoretical Concepts } 
\label{theo_conc}
\begin{figure}[!htb]
\centering
\includegraphics[scale=1.2]{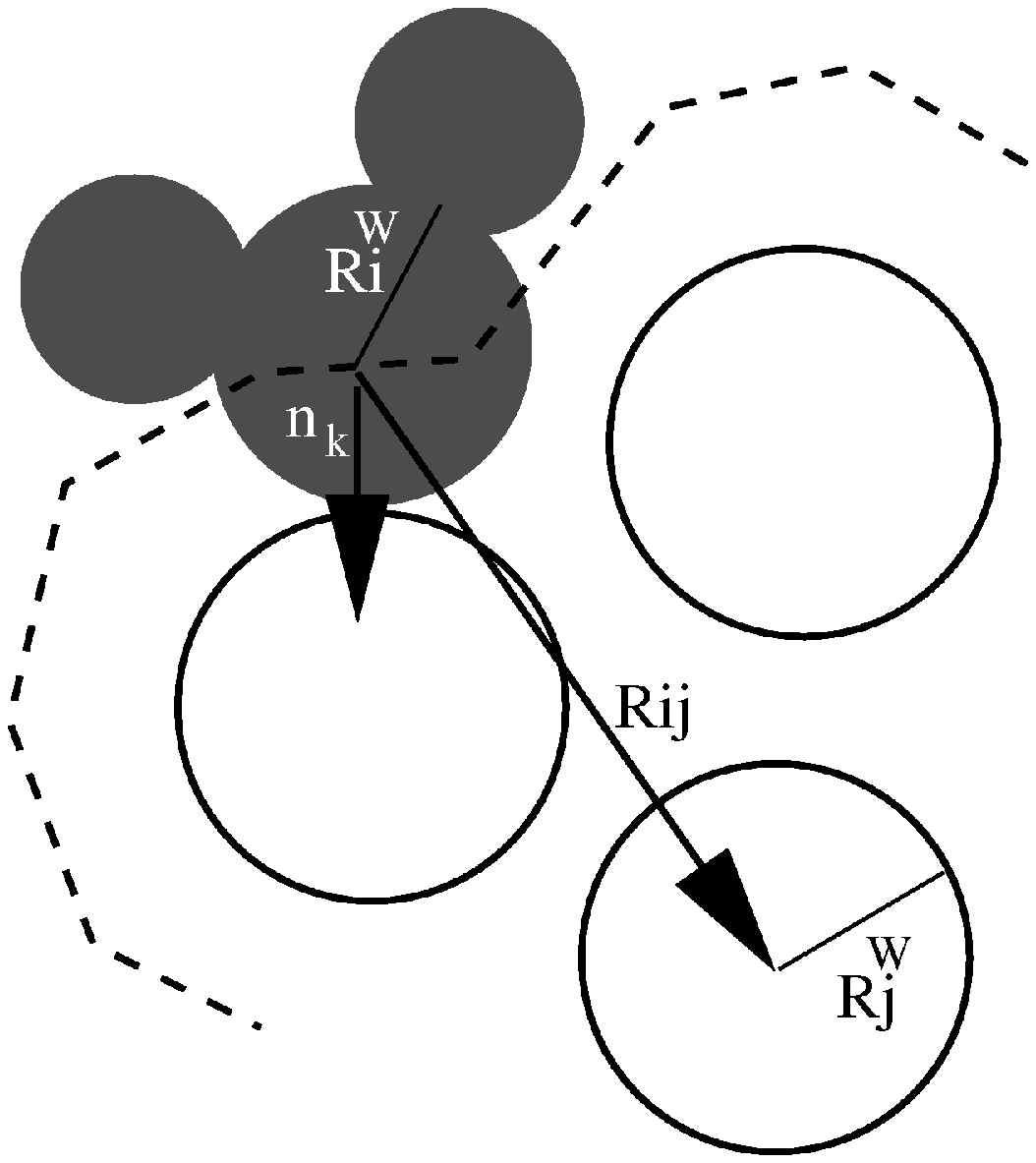}
\caption[  Graphical representation of the geometrical elements
           needed for computing the dispersion energy according to 
           Floris et al. \cite{floris}. The solvent molecule is given 
           in grey, while a few atoms of the solute are shown in white. 
           The solute-solvent boundary is indicated as dashed line.
           For details of the algorithm please refer to the explanation
           given in section  \ref{theo_conc}.                            ]
           { \label{fig1} }
\end{figure}

We use the following decomposition of the solvation free energy
\begin{equation}
\label{eq1}
  \Delta G^{solv} = \Delta G^{pol} + \Delta G^{cav} + \Delta G^{disp}  
\end{equation}
where the individual terms represent polarization, cavitation and 
dispersion contributions. Explicit consideration of repulsion is not
necessary as the cavitation term includes these interactions. PB/BEM 
methodology is used for $\Delta G^{pol}$ at the boundary specification 
described 
\textcolor{red}{previously} 
\cite{parimal}. The cavitation term is expressed via the revised Pierotti 
approximation \cite{hoefi,mahajan}
\textcolor{red}{(rPA), which is based on the Scaled Particle Theory 
                \cite{reiss,pierotti}. The major advantage with this revised
                approximation is a transformation property involving
                the solvent excluded volume. Hence after having identified the
                basic rPA-coefficients from free energy calculations the
                rPA-formula may be applied to any solute regardless of its
                particular shape or size \cite{hoefi}.
               }
$\Delta G^{disp}$ is computed 
from the Caillet-Claverie formula \cite{caillet,claverie} projected onto the 
boundary elements as suggested by Floris et al. \cite{floris}
\begin{equation}
\label{eq2}
  \Delta G^{disp} = \sum\limits_i^I 
                       \rho^{slv} \omega_i
                    \sum\limits_j^J
                    \sum\limits_k^K
                    \underbrace{  -0.214 \kappa_i \kappa_j 
                                  \frac{ 64 (R_i^W)^3 (R_j^W)^3 }
                                       { {R_{ij}}^6 }
                               }_{Caillet-Claverie}
                    \frac{1}{3} 
                    \left( 
                       \vec{R}_{ij} \cdot \vec{n}_k 
                    \right)
                    \Delta \sigma_k
\end{equation}
where the first sum is over different atom types, $i$, composing one molecule
of solvent, the second sum is over all solute atoms, $j$, and the sum over 
$k$ is over all surface elements resulting from an expansion of the 
molecular surface by the dimension of radius $R_i^W$ of a particular solvent 
atom, see Figure \ref{fig1} for a graphical representation. Here solvent atoms 
are shown in grey and solute atoms are represented as white circles. The scheme
corresponds to one particular choice of $i$. For example, if the solvent 
molecule in Figure \ref{fig1} is water, then the scenery depicts the first of 
two possibilities where $i$ refers to the oxygen atom. After $i$ is set, all 
atom radii of the solute are increased by the amount of the atomic radius of 
oxygen and the molecular surface (dashed line in Figure \ref{fig1}) is 
reconstructed. Next the inner double sum is carried out where $J$ is the total 
number of solute atoms and $K$ is the total number of BEs forming the 
interface. Note that index $j$ serves for a double purpose, looping over all 
solute atoms as well as defining the type of atomic radius to use. At every 
combination $j,k$ of solute atoms with BEs, the expression emphasized by the 
curly bracket in eq. \ref{eq2} must be evaluated. Here $\kappa_i$ and 
$\kappa_j$ are dispersion coefficients and $R_i^W$, $R_j^W$ are atomic radii, 
all of them determined empirically by Caillet-Claverie \cite{caillet,claverie}.
The corresponding values are summarized in Table \ref{table1}.  $R_{ij}$ is 
the distance between the center of some BE, $k$, and the center of a solute 
atom, $j$. After the expression in the curly bracket of eq. \ref{eq2} has been 
evaluated it must be multiplied with a scalar product between the vectors 
$\vec{R}_{ij}$ and $\vec{n}_k$, the inwards pointing normal vector 
corresponding to the $k^{th}$ BE. The remainder of eq. \ref{eq2} is  
multiplication with a constant factor $\frac{1}{3}$ and multiplication with
$\Delta \sigma_k$, the partial area of the BE, $k$. After all possible 
combinations $j,k$ have been considered, the procedure is repeated with
an incremented $i$, now referring to the H-atom, the second type of atom in 
a molecule of water. The molecular surface is recomputed, extended
by the dimension of the atomic radius of hydrogen, and the entire inner
double sum will be repeated as outlined for the case of oxygen.
However, since both H-atoms in the solvent molecule are identical this step 
needs to be done only once and $\omega_i$, the number of occurrences of a 
particular atom type $i$, will take care of the rest (in the case of water
$\omega_1=1$ for oxygen and $\omega_2=2$ for hydrogen). Finally, $\rho^{slv}$ 
in eq. \ref{eq2} represents the number density of the solvent. We have 
restricted the approach to just the 6$^{th}$-order term in the expression 
derived by Floris et al. \cite{floris}. Note that we consider molecular 
surfaces as defined by Connolly \cite{connolly}. The partial term listed after 
the curly bracket in eq. \ref{eq2} is the actual consequence of mapping the 
classical pair interaction terms onto a boundary surface \cite{floris}. The 
partial expression enclosed in the curly bracket can be substituted with any 
other classic pair potential, for example using AMBER style of dispersion 
\cite{amber},
\begin{equation}
\label{eq3}
  \Delta G^{disp} = \sum\limits_i^I  
                       \rho^{slv} \omega_i
                    \sum\limits_j^J
                    \sum\limits_k^K
                    \underbrace{  -2 \sqrt{ \epsilon_i \epsilon_j }
                                  \left(
                                     \frac{ R_i^W + R_j^W }
                                          { R_{ij} }
                                  \right)^6
                               }_{AMBER}
                    \frac{1}{3}
                    \left(
                       \vec{R}_{ij} \cdot \vec{n}_k
                    \right)
                    \Delta \sigma_k
\end{equation}
with similar meanings of the variables used above and $\epsilon_i$ being
the van der Waals well depth corresponding to homogeneous pair interaction
of atoms of type $i$.

\subsection{ Model Calibration } 
\label{model_calib}
The algorithm covering computation of dispersion is implemented in the PB/BEM 
program POLCH \cite{hoefi1} (serial version). Proper functionality was tested 
by comparing dispersion results of 4 sample molecules, methane, propane, 
iso-butane and methyl-indole, against results obtained from GAUSSIAN-98 
\cite{g98} (PCM model of water at user defined geometries). Deviations were on 
the order of $\pm$ 1.8 \% of the G98 value, so the procedure is assumed to 
work correctly. The small variations are the result of employing a different 
molecular surface program in PB/BEM \cite{vorobjev}. Next the structures of 
amino acid side-chain analogues are derived from standard AMBER pdb \cite{pdb} 
geometries by making the C$_\alpha$-atom a hydrogen atom, adjusting the C-H 
bond length and deleting the rest of the pdb structure except the actual side 
chain of interest. In a similar process, zwitterionic forms of each type of 
amino acid are constructed. PB/BEM calculations are carried out and net 
solvation free energies for solvent water are stored. A comparison is
made against the experimental values listed in \cite{chang} as well as 
results obtained from the PCM model in GAUSSIAN-03 \cite{g03}. AMBER default 
charges and AMBER van der Waals radii increased by a multiplicative factor 
of 1.07 are used throughout \cite{parimal}. Initial deviation from the 
reference set is successively improved by introducing a uniform scaling 
factor to the dispersion coefficients $\kappa_i$ of eq. \ref{eq2}. The 
optimal choice of this dispersion scaling factor is identified from the 
minimum mean deviation against the reference data set. The initially
derived optimal scaling factor is applied to the zwitterionic series, a 
subset of molecules for which experimental values have been compiled \cite{li},
and a set of 180 dipeptide conformations studied previously. When new 
molecules are parameterized, we use ANTECHAMBER from AMBER-8 and 
RESP charges based on MP2/6-31g* grids of electrostatic potentials 
\cite{cornell}. Molecular geometries are optimized in a two-step procedure, 
at first at B3LYP/3-21g* and then at MP2/6-31g* level of theory and only the 
final optimized structure becomes subject to the RESP calculation. 

Extensions are pursued in two directions. First, the PB/BEM approach is used 
with solvents other than water, and the question is raised whether the 
optimized scaling factor for dispersion in water is of a universal nature or 
needs to be re-adjusted for each other type of solvent considered. Secondly,
we tested the introduced change when the Caillet-Claverie specific formalism 
of dispersion  treatment is changed to AMBER-style dispersion as indicated in 
eqs.  \ref{eq2} and \ref{eq3}.

\subsection{ Study of Size- and Charge Dependence }
Crystal structures of 10 proteins of increasing size are obtained from the 
Protein Data Bank \cite{pdb}. The actual download site is the repository 
PDB-REPRDB \cite{pdb1}. Structures are purified and processed as described 
previously \cite{parimal2}. The PDB codes together with a characterization of 
main structural features of the selected test proteins are summarized in Table 
\ref{table2}. Two types of calculations are carried out using the 
semi-empirical model PM5 \cite{stewart} and the fast multipole moment (FMM) 
method \cite{white}. A single point vacuum energy calculation is followed
by a single point energy calculation including the COSMO model \cite{klamt}
for consideration of solvent water. The difference between the two types
of single point energies should provide us with an estimate of the solvation 
free energy. Furthermore, the finally computed set of atomic partial charges 
is extracted from the PM5-calculation and feeded into the PB/BEM model to 
substitute standard AMBER partial charges. In this way we can examine the 
dependence on a chosen charge model as well as compare classic with 
semi-empirical QM approaches to the solvation free energy.

\textcolor{red}{
                \subsection{ Computational Aspects }
                The sample set of 10 proteins listed in Table \ref{table2} is 
                analyzed with respect to computational performance regarding
                the calculation of the dispersion term as defined in eq. 
                \ref{eq3}. It is important to note that for this particular 
                task the surface resolution into BEs may be lowered to levels 
                where the average size of the BEs becomes $\approx$ 0.45 
                \AA$^2$. CPU times for the two steps, ie creation and 
                processing of the surface and evaluation of the expression 
                for $\Delta G^{disp}$ are recorded and summarized in 
                Table XII of the Supplementary Material. As can be 
                seen clearly from these data, the major rate-limiting step
                is the production of the surface, which can reach levels
                of up to 20 \% of the total computation time. Evaluation
                of the dispersion term itself is of negligible computational
                cost. Since the surface used for the polarization term 
                is defined according to Connolly (see section \ref{theo_conc}),
                we could not use this molecular surface directly for a 
                SASA-based alternative treatment of the non-polar
                contributions. Rather we had to compute a SASA from
                scratch too, and were facing identical computational 
                constraints as seen with the approach chosen here. 
               }

\section{ Results }
\subsection{ A universal scaling factor applied to Caillet-Claverie 
             dispersion coefficients leads to good overall agreement with
             experimental solvation free energies of amino acid side-chain 
             analogues in water                                            }
\label{scalf_0.70}
\begin{figure}[!htb]
\centering
\includegraphics[scale=1.0]{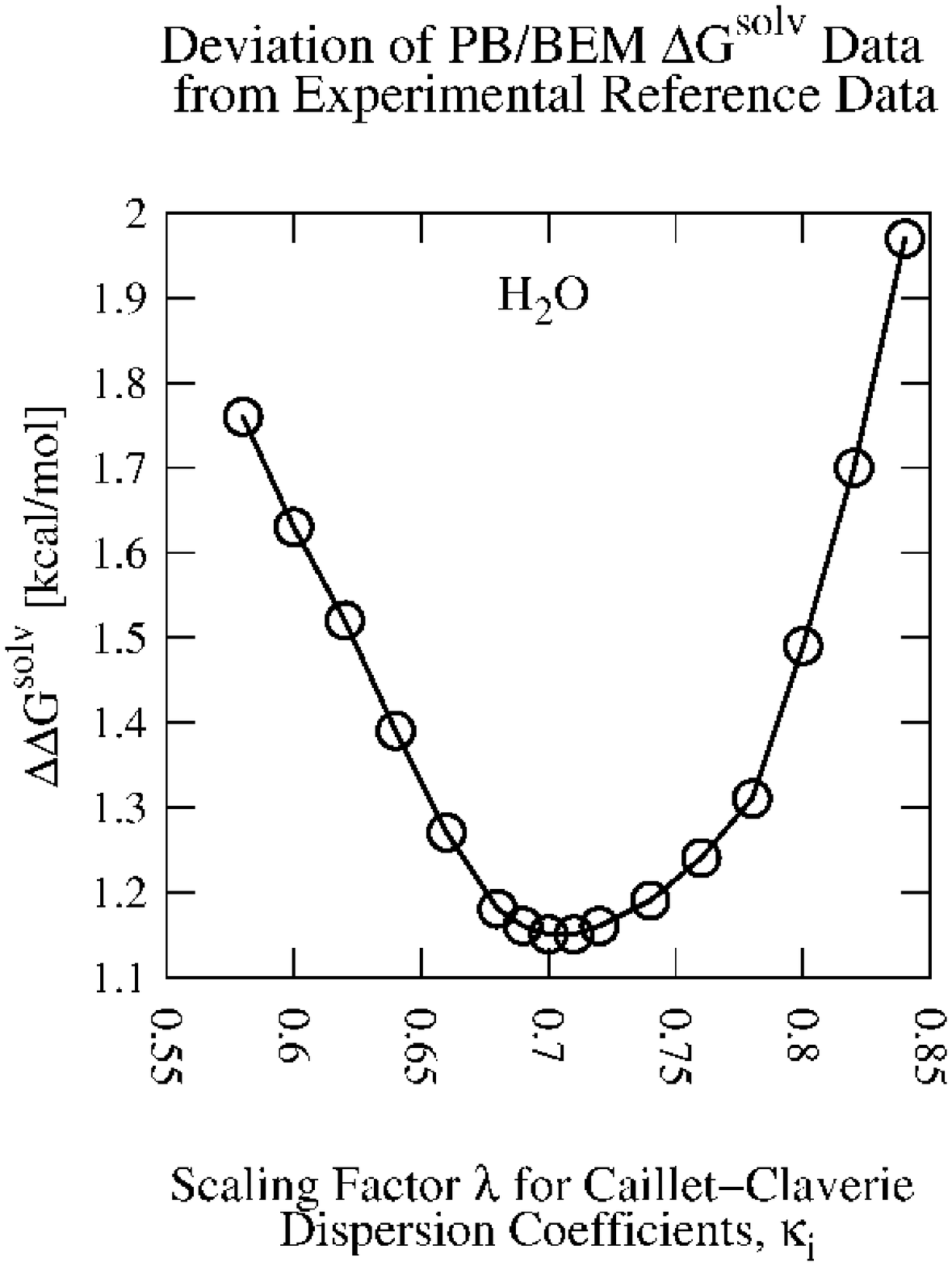}
\caption[  Deviation of the PB/BEM solvation free energies 
           $\Delta G^{solv}$ from experimental values tabulated in
           \cite{chang} as a function of $\lambda$, a scaling factor 
           uniformly applied to all Caillet-Claverie 
           \cite{caillet,claverie} dispersion coefficients 
           $\kappa_i$ (see eq. \ref{eq2}).                            ]
        { \label{fig2} }
\end{figure}

Since our main focus is on proteins, our first goal is to optimize our 
approach for proteins in aqueous solution. We can resort to the 
experimental data for amino acid side-chain analogues (see \cite{chang} 
and references therein). At first we seek maximum degree of agreement 
between experimental and PB/BEM values of the solvation free energy,
$\Delta G^{solv}$, by multiplying a scaling factor, $\lambda$, to the 
Caillet-Claverie dispersion \cite{caillet,claverie} coefficients, $\kappa_i$. 
The remaining terms in eq. \ref{eq1} are computed at the optimized conditions 
reported previously \cite{parimal,mahajan}. We define a global deviation from 
the experimental data by
\begin{equation}
\label{eq4}
  \Delta \Delta G^{solv} = \frac{1}{13}
                  \sum\limits_{i=1}^{13}
                  \sqrt{ \left(
                            \Delta G^{solv,Exp}_i 
                            - 
                            \Delta G^{solv,PB/BEM}_{i,\lambda}
                         \right)^2                                  
                       }
\end{equation}
where $i$ refers to a particular type of amino acid side-chain analogue 
included in the reference set of experimental values and $\lambda$ is the 
introduced scaling factor applied to the Caillet-Claverie dispersion 
\cite{caillet,claverie} coefficients. The trend of $\Delta \Delta G^{solv}$ 
for different choices of $\lambda$ is shown in Figure \ref{fig2}. As becomes 
clear from Figure \ref{fig2}, a scaling factor of 0.70 establishes the best 
match to the experimental data. A detailed comparison of individual amino 
acid side-chain analogues at this optimum value is given in Table 
\ref{table3}. We achieve a mean unsigned error of 1.15  $\frac{kcal}{mol}$, 
hence come close to the accuracy reported recently by Chang et al. 
\cite{chang}, a study that agreed very well with earlier calculations carried 
out by Shirts et al. \cite{shirts} and MacCallum et al. \cite{maccallum}.
\textcolor{red}{ Several computed solvation free energies in Table \ref{table3}
                 still show significant deviation from the experimental value,
                 e.g. p-cresol and methanethiol. A comparison to results 
                 with a simple SASA-based model is included in the 
                 Supplementary Material (Table XI). This comparison reveals 
                 a certain improvement for the most critical components, but 
                 no indication of a general amelioration of the situation.
                 The somewhat special character of methanethiol has been 
                 noticed before \cite{simonson}.
               }

\subsection{ Component-wise juxtaposition of PB/BEM and PCM approaches   
             reveals a difference in individual contributions but 
             similarity in net effects                                  }
As interesting as total solvation free energies are the constituting 
partial terms and how they compare to their analogous counter parts in 
a high-level QM model such as PCM. We therefore studied all amino acid 
side-chain analogues with PCM \cite{tomasi} calculations at the Becke-98 
\cite{becke} level of density functional theory (DFT) using the high-quality 
basis set of Sadlej \cite{sadlej} and program GAUSSIAN-03 \cite{g03}. A
summary of these data is given in Table \ref{table4}. Since in PB/BEM we do 
not consider repulsion explicitly, the PB/BEM dispersion term is compared to 
the sum of $\Delta G^{disp}$ and $\Delta G^{rep}$ of PCM. It becomes clear
from Table \ref{table4} that there is rather general agreement in polarization 
terms but sizeable divergence in the apolar terms. However, the sum of
all apolar terms, ie. $\Delta G^{cav}$ and $\Delta G^{disp}$, appears to be 
again in good agreement when comparing PB/BEM with PCM. The reason for the
difference in the apolar terms is largely due to a different cavitation 
formalism used in PB/BEM, which we currently believe to represent a very 
good approximation to this term \cite{mahajan}.

\subsection{ The identified scaling factor of 0.70 applied to 
             Caillet-Claverie dispersion coefficients yields good 
             quality estimates of the solvation free energy in water for 
             many molecules                                              } 
In order to test the PB/BEM approach further we used the initially
determined scaling factor for dispersion coefficients of 0.70 to compute
water solvation free energies of a series of other molecules. The procedure
for obtaining atomic partial charges is described in section 
\ref{model_calib}. It is important to note that the electron density used 
for RESP fitting must be of MP2/6-31G* quality to achieve maximum degree
of compatibility to standard AMBER charges, which have been found to 
mimic high quality calculations very well \cite{parimal}. Experimental
reference values have been obtained from the extensive compilation by 
Li et al. \cite{li}. The data comprising 18 arbitrarily selected molecules 
are summarized in Table \ref{table5}. The mean unsigned error of 1.18 
$\frac{kcal}{mol}$ for this set of molecules comes close to PCM quality and 
must be considered very satisfactory again. 

Another class of molecules we looked into are amino acids in their 
zwitterionic form, where due to the charges at the amino/carboxy groups the 
net solvation free energies become larger by about an order of magnitude. A 
comparison against the recently reported data by Chang et al. \cite{chang} is 
given in Table \ref{table6}. The degree of agreement is still considerably 
high and there is no obvious indication of a systematic deviation. 

A final comparison is made against a series of 180 molecules that has 
been used in a previous study \cite{parimal}. These structures include all
20 types of naturally occurring amino acids in 9 different conformations
(zwitterionic forms assumed). The set of dipeptides has been subjected 
to PCM \cite{tomasi} calculations at the Becke-98 DFT level \cite{becke} 
using Sadlej's basis set \cite{sadlej}. Average net solvation free energies 
are formed from all 9 different conformations per type of amino acid (or the 
number of available reference calculations) and the results are presented in 
Table I in the supplementary material. Considering the variation with respect 
to conformational flexibility the match must still be considered to be 
reasonably good. It is interesting to note that the variability of the 
dispersion contributions alone, considered isolated per se as a function of 
conformational flexibility is much less pronounced than what we see for the 
net solvation (see Table II of the supplementary material).

\subsection{ The scaling factor of 0.70 applied to Caillet-Claverie 
             dispersion coefficients in the case of water is not of a
             universal nature but must be re-optimized for any other type 
             of solvent.                                                   }
\begin{figure}[!htb]
\centering
\includegraphics[scale=1.0]{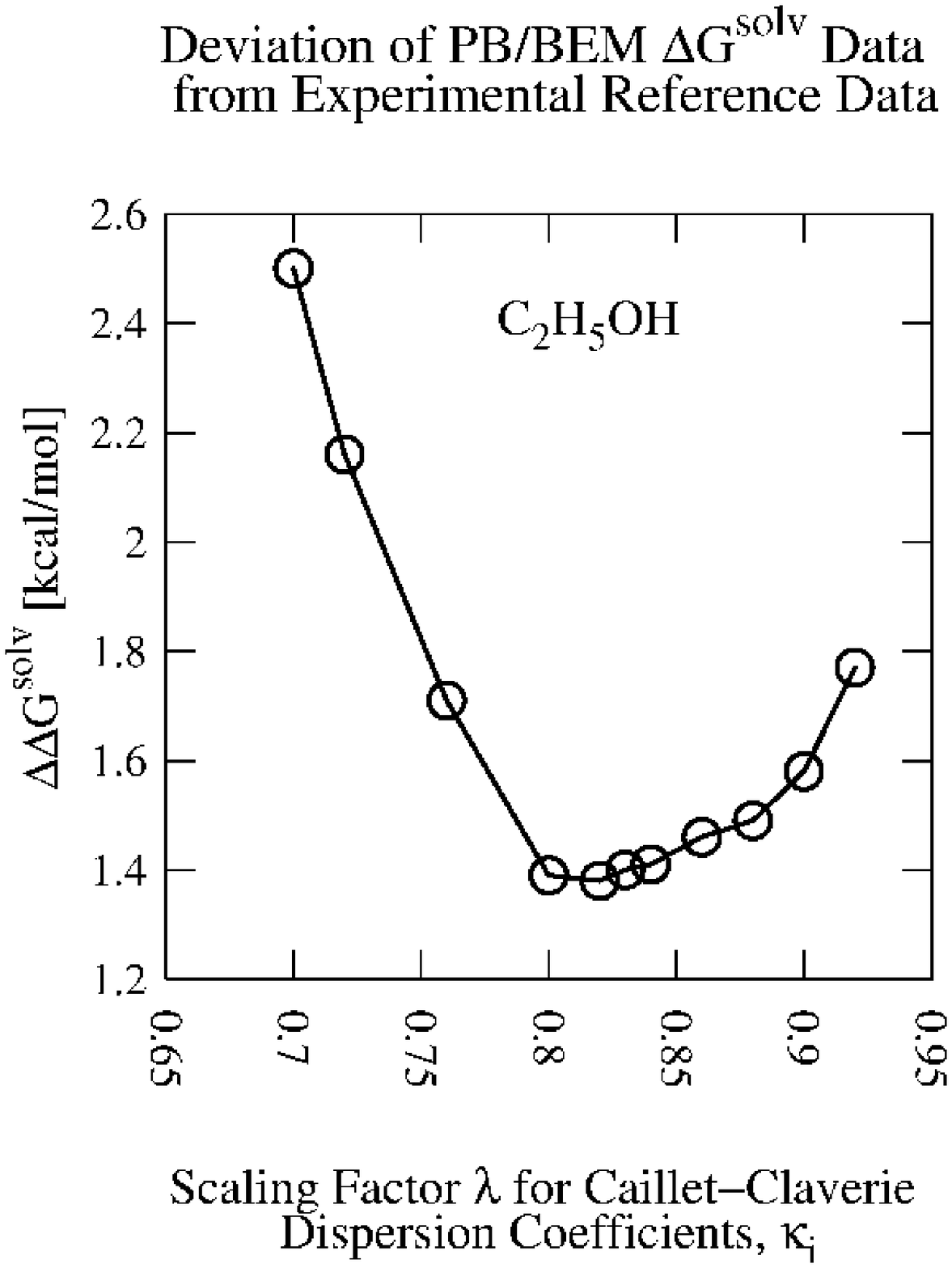}
\caption[  Ethanol: Deviation of the PB/BEM solvation free energies
           $\Delta G^{solv}$ from experimental values tabulated in
           \cite{li} as a function of $\lambda$, a scaling factor 
           uniformly applied to all Caillet-Claverie 
           \cite{caillet,claverie} dispersion coefficients 
           $\kappa_i$ (see eq. \ref{eq2}).                            ]
        { \label{fig3} }
\end{figure}

\begin{figure}[!htb]
\centering
\includegraphics[scale=1.0]{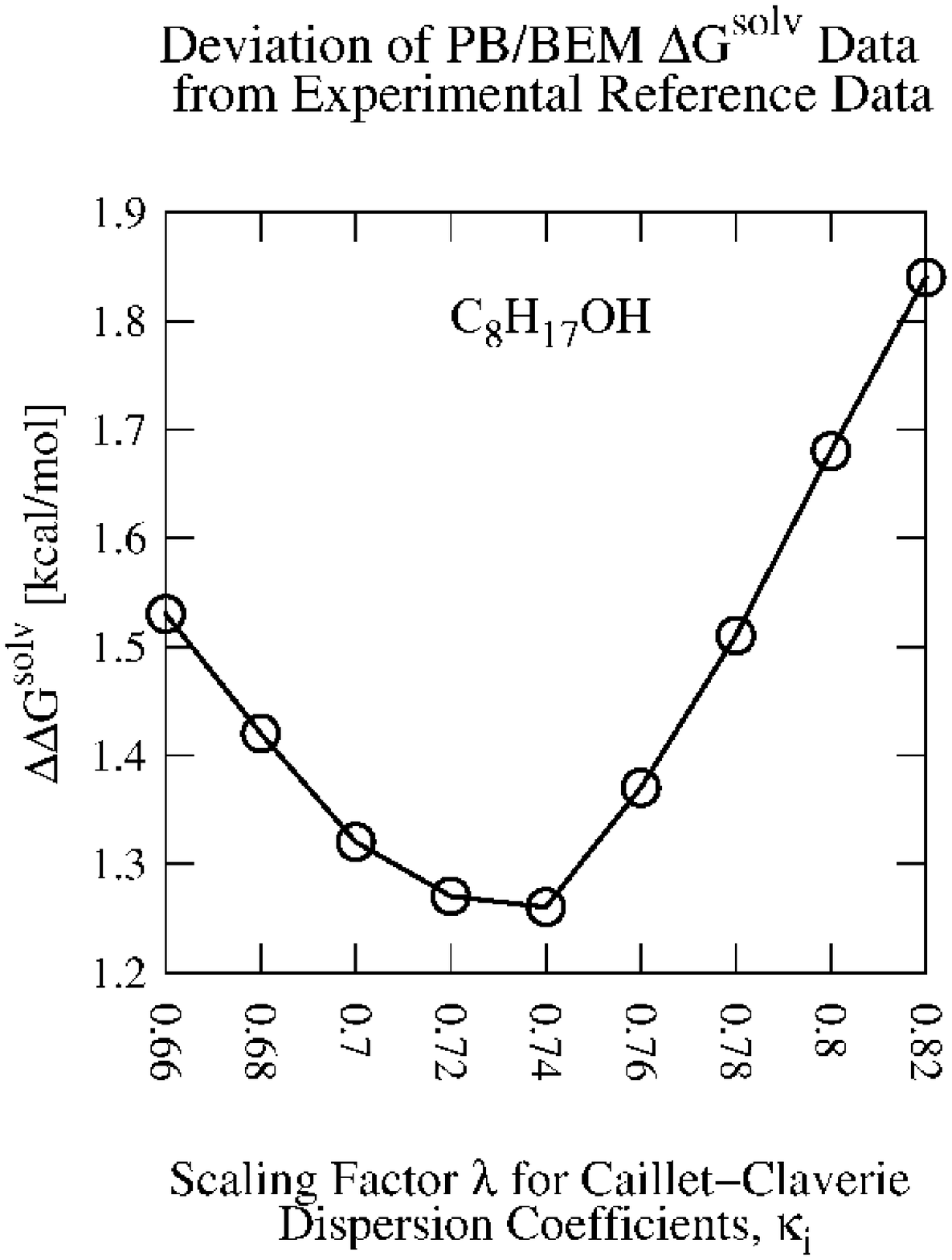}
\caption[  n-Octanol: Deviation of the PB/BEM solvation free energies
           $\Delta G^{solv}$ from experimental values tabulated in
           \cite{li} as a function of $\lambda$, a scaling factor 
           uniformly applied to all Caillet-Claverie 
           \cite{caillet,claverie} dispersion coefficients 
           $\kappa_i$ (see eq. \ref{eq2}).                            ]
        { \label{fig4} }
\end{figure}

An important aspect of the PB/BEM approach is how the identified scaling
factor for Caillet-Claverie dispersion coefficients --- 0.70 in the case
of water --- translates into other situations of non-aqueous solvation. We 
have therefore repeated the studies for identifying optimal boundaries 
\cite{parimal} for solvents methanol, ethanol and n-octanol. Again we consider 
PCM cavities of the set of 180 dipeptide structures as reference systems and 
search for the best match in volumes and surfaces dependent on slightly 
enlarged or shrinked standard AMBER van der Waals radii. We again employ the 
molecular surface program SIMS \cite{vorobjev}. Detailed material of this fit 
is included in the supplementary material (Tables III-VIII and Figures I-VI).
We find to have to marginally increase AMBER van der Waals radii by factors 
of 1.06 in solvents methanol and ethanol and 1.05 in solvent n-octanol.
Based on these conditions for proper locations of the solute-solvent
interface we then repeat the search for appropriate scaling factors of 
dispersion coefficients that result in close agreement to experimental
solvation free energies (see section \ref{scalf_0.70}). Results are presented 
in Figures \ref{fig3} and \ref{fig4}. It becomes clear that the factor of 0.70,
optimal for water, is not universally applicable. Rather, we find for ethanol 
0.82 and for n-octanol 0.74 to be the optimal choices. A detailed comparison 
against experimental values at optimized conditions is given in Tables 
\ref{table7} and \ref{table8}. We achieve mean unsigned errors of 1.38 
$\frac{kcal}{mol}$ for ethanol and 1.27 $\frac{kcal}{mol}$ for n-octanol. 
Cavitation terms of similar quality to the ones presented in \cite{mahajan}, 
which are needed in PB/BEM, are available for methanol and ethanol 
(unpublished work in progress) or obtained from \cite{hoefi2}. Unfortunately,
we cannot do the calculations for methanol because of the lack of experimental 
values and the non-systematic trend in dispersion scaling factors of the other 
alcoholic solvents. All optimized parameter sets for the various types of 
solvents are summarized in Table \ref{table9}.

\textcolor{red}{
                \subsection{ Switching from Caillet-Claverie-style of 
                             dispersion to AMBER-style requires a 
                             re-adjustment of scaling factors.        }
                An obvious question is how the described approach will change 
                when substituting the Caillet-Claverie formalism with the 
                corresponding AMBER-dispersion formula, ie replacing
                eq. \ref{eq2} with eq. \ref{eq3}. We therefore implemented a 
                variant where we use eq. \ref{eq3} together with standard 
                AMBER van der Waals radii (slightly increased as done for the 
                definition of the boundary and indicated in table \ref{table9})
                and standard AMBER van der Waals potential well depths. 
                Similar to the Caillet-Claverie treatment we find that 
                a uniform scaling factor, $\lambda$, applied to the AMBER 
                van der Waals potential well depths, $\epsilon_i$, is 
                sufficient to lead to good agreement with experimental data. 
                An identical strategy to the one presented in section 
                \ref{scalf_0.70} for determination of appropriate values
                of $\lambda$ may be applied. The optimal choice of $\lambda$ 
                turns out to be 0.76 for solvent water as indicated in Figure 
                IX of the Supplementary Material. Corresponding detailed data 
                is shown in Table \ref{table10}. The mean unsigned error 
                amounts to 1.01 $\frac{kcal}{mol}$ at optimized conditions. 
                While in the case of water similar scaling factors are 
                obtained for Caillet-Claverie as well as AMBER type of 
                dispersion, for the remaining types of solvents a less 
                coherent picture arises (see Table \ref{table9}). 
                Identification of scaling factors for solvents ethanol 
                ($\lambda$=0.94) and n-octanol ($\lambda$=2.60) is shown       
                in Figures X and XI of the Supplementary Material and 
                corresponding detailed data listed in Tables IX and X of the 
                Supplementary Material. Mean unsigned errors are 1.21 
                $\frac{kcal}{mol}$ for ethanol and 1.00 $\frac{kcal}{mol}$ for 
                n-octanol respectively. Either approach is competitive and
                comes with its own merits. Caillet-Claverie coefficients are 
                more general and specific to chemical elements only, hence no 
                distinction between for example sp3-C atoms and sp2-C atoms 
                needs to be made. Employment of AMBER parameters on the
                other hand appears to be straightforward in the present 
                context since the geometry of the boundary is already based 
                on AMBER van der Waals radii.
               }

\subsection{ Replacement of static AMBER partial charges with semi-empirical
             PM5 charges introduces a rise in solvation free energies by about
             20 \% of the classic result regardless of the size or total 
             charge state of the system.                                      } 
\begin{figure}[!htb]
\centering
\includegraphics[scale=1.4]{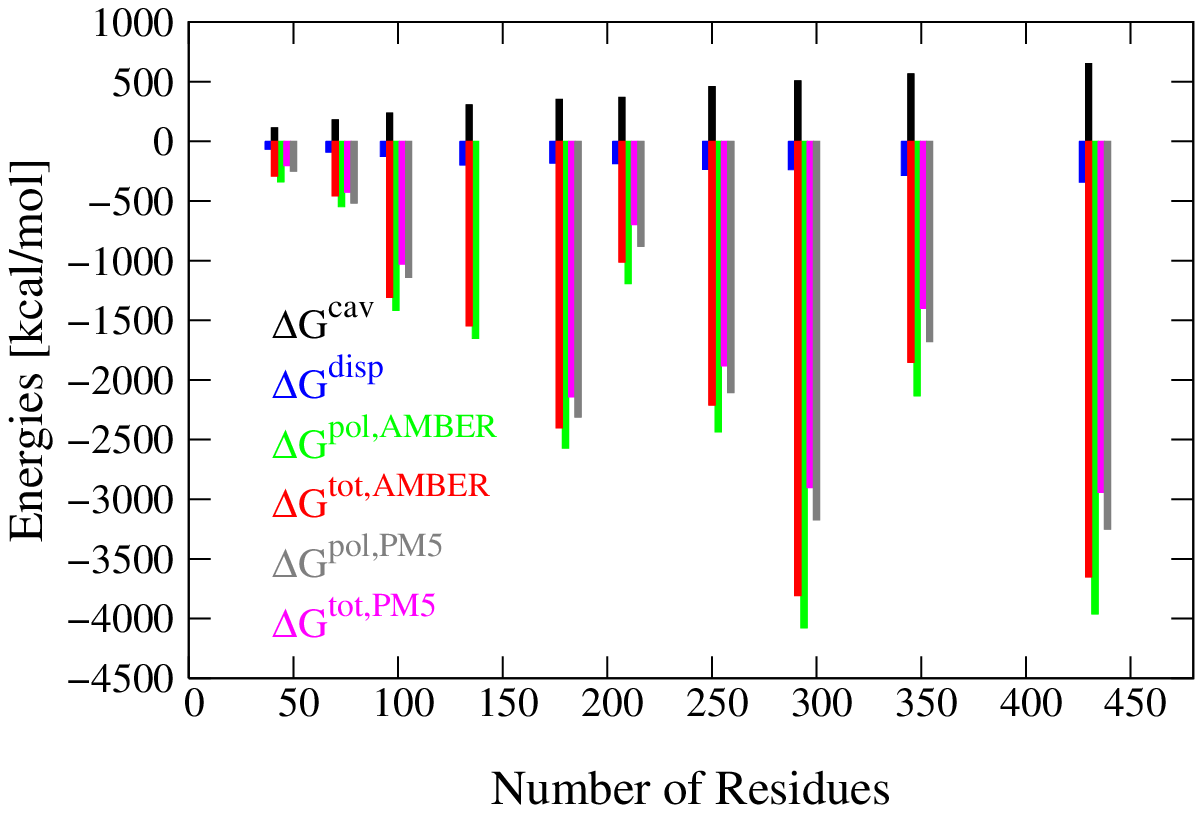}
\caption[  Classic versus semi-empirical charge assignments to atoms
           of proteins of various size used in PB/BEM calculations.
           AMBER standard charges are used in the classic approach,
           while PM5 charges are taken in the semi-empirical variant.
           Partial terms are colour-coded as indicated and also 
           listed in Table \ref{table11}.                             ]
        { \label{fig5} }
\end{figure}

A series of proteins of different size, shape and total net charge (see
Table \ref{table2}) is computed within the PB/BEM approach at optimized
conditions for aqueous solvation, that is using a Caillet-Claverie dispersion 
coefficient scaling factor of 0.70, slightly increased AMBER van der Waals 
radii by a factor of 1.07 and standard AMBER partial charges. In addition 
to this classic approximation we also carry out semi-empirical QM calculations
with the help of program LocalSCF \cite{anikin} using the PM5 model. From
the semi-empirical calculation we extract atomic partial charges and use these
instead of AMBER partial charges within the PB/BEM approach. Results of these
calculations are presented in Table \ref{table11} and Figure \ref{fig5}. In
general one can observe a rather constant 
\textcolor{red}{ change of about }
20 \% of the classic AMBER based $\Delta G^{solv}$ estimate when switching to 
PM5 charges. 
This is independent of the size, shape or net charge of the system (compare
red bars with purple bars in Figure \ref{fig5}). The polarization term 
constitutes the major contribution but apolar terms are far from negligible
(compare magnitude of blue and black bars to green and grey bars in Figure 
\ref{fig5}). When using the COSMO approximation within the semi-empirical    
method and deriving solvation free energies from that we get entirely 
uncorrelated results for the solvation free energy, $\Delta G^{solv}$ (see
9$^{th}$ column in Table \ref{table11}). It is important to note that the
surface to volume ratio drops to a value around 0.25 with increasing protein
size, whereas typical values in the range of 0.80 to 1.0 are maintained in 
the initial calibration phase, hence care must be taken with large scale 
extrapolations from small molecular reference data.

\section{ Discussion }
Motivated by our recent high-performance implementation of Poisson-Boltzmann
calculations \cite{hoefi1} we now complement this approach with a systematic
inclusion of apolar effects. In particular the important dispersion
contribution is introduced and fine-tuned against available experimental data.
This is based on physics-based terms, that have long been considered in a 
similar fashion within QM models \cite{tomasi}. The resulting model is applied 
to a series of protein structures, and size and charge effects are examined. 

Direct assessment of the predictive quality of the PB/BEM approach after 
calibration has revealed rather good performance indicators for PB/BEM.     
This was based on suggested scaling factors applied to Caillet-Claverie 
dispersion coefficients. Since the original aim of Caillet-Claverie was to 
explain crystal data, we would expect a need for re-adjustment in this present 
implementation. Moreover, since the boundary and the rest of the PB/BEM model 
is based on AMBER parameterization it does not come as a surprise that one has 
to adjust a non-related second set of van der Waals parameters in order to 
achieve general agreement to a reference data set. Related to this point it 
seems particularly encouraging that when replacing the scaled Caillet-Claverie 
part with standard AMBER-dispersion terms for water no further refinement is 
necessary and similar levels of precision are established automatically. In 
the case of water, this brings in a second advantage. Because the employed 
TIP3P model assigns van der Waals radii of zero to the H-atoms, so the 
effective sum over $i$ in eq. \ref{eq3} may be truncated already after the 
oxygen atom. The second cycle considering H-atoms in water would add only 
zeros any more.    

A somewhat critical issue is the determination of missing parameters
or the estimation of solvent probe sphere radii for different types of
solvents. In this present work we found it convenient to make use of electron 
density grids and corresponding iso-density thresholds to define the boundary 
of molecules. For example to determine the probe sphere radius of methanol we 
compute the volume of a single molecule of methanol up to an electron density 
threshold of 0.0055 a.u. and derive an effective radius assuming spherical 
relationships. The same threshold criterion is applied to all other solvents
leading to the data summarized in Table \ref{table9}. Electron grids are
based on B98/Sadlej calculations. Similarly we determine atomic van der Waals
radii for Cl- or Br-containing substances from iso-density considerations. 
However in these latter cases the threshold criterion is adapted to a level
that re-produces proper dimensions of well-known types, ie neighboring
C-, O-, N-atoms and at this level the unknown radius is determined. In the 
case of n-octanol the assumption of spherical geometry is certainly not
justified. On the other hand the concept of an over-rolling probe sphere
representing approaching solvent molecules will remain a hypothetical model 
construct anyway. Complying with this model construct it may be argued that 
over time the average of approaching solvent molecules will hit the solute 
with all parts (head, tail or body regions of the solvent molecule) equally 
often and thus the idealized spherical probe is not entirely unreasonable.

Another interesting aspect is the fact that the present PB/BEM approach
is all based on molecular surfaces rather than SASAs. This is of technical
interest and the consequence of that is a greatly reduced sensitivity to 
actual probe sphere dimensions. A graphical explanation is given in the 
supplementary material (Figure VII). While SASA based surfaces would 
see significant changes when probe spheres are slightly modified (blue
sphere replaced by red sphere in Figure VII of the supplementary material)
the molecular surface itself faces only a minor change in the reentrant
domain (green layer indicated in Figure VII of the supplementary material).

Large scale extrapolations resulting from a calibration process done with 
small sized reference structures have to be taken with care. Because of the 
drop of surface to volume ratios the most important requirement for such a 
strategy is to have the individual terms properly analyzed whether they 
scale with the volume, or the surface. For PB/BEM the question reduces to the 
cavitation term, since the remainder is mainly a function of Coulombic 
interactions. As that particular aspect has been carefully analyzed in
 previous studies \cite{hoefi} we are confident that a large-scale 
extrapolation actually works in the way suggested in eq. \ref{eq1}. 

A final remark may be relevant with regard to the discrepancy seen in using 
classic AMBER partial charges versus semi-empirical PM5 charges. Intuitively, 
one is tempted to believe stronger in the PM5 results. There might however
also be a small drift in energies introduced by PM5/PM3 models as has been 
observed within an independent series of single point calculations (see
supplementary material Figure VIII).

\section{ Conclusion }
Consideration of dispersion effects within a physics-based continuum solvation
model significantly improves accuracy and general applicability of such
an approach. The proposed method follows a proven concept \cite{floris}
and is easily implemented into existing models. Generalization to different
treatments of dispersion as well as extension to non-aqueous solvents is
straightforward.

\section{ Acknowledgements }
This work was supported by the National Institutes of Health Grant GM62838. 
We thank FQS Poland and the FUJITSU Group, for kindly providing a 
temporary test version of program LocalSCF \cite{anikin}.
\pagebreak

\clearpage
\pagebreak

\begin{table}[!htb]
\begin{center}
\caption[]{\label{table1} Summary of the data used for Caillet-Claverie
           style of dispersion treatment as outlined in eq. \ref{eq2}. }
\vspace{0.5cm}
\begin{tabular}{cccccccccccc}
\\ \hline\hline \\
\multicolumn{12}{c}{Caillet-Claverie Dispersion Coefficients, $\kappa$,
                    and Atomic Radii, $R^{W}$, in \AA 
                    \cite{caillet,claverie}                           }  \\
\\ \hline \\
  $\kappa_H$    & $\kappa_C$    & $\kappa_N$    & $\kappa_O$   & 
  $\kappa_F$    & $\kappa_{Na}$ & $\kappa_{P}$  & $\kappa_S$   & 
  $\kappa_{Cl}$ & $\kappa_K$    & $\kappa_{Br}$ & $\kappa_J$   \\
  1.00          & 1.00          & 1.18          & 1.36         & 
  1.50          & 1.40          & 2.10          & 2.40         & 
  2.10          & 2.90          & 2.40          & 3.20         \\
\\    
  $R_H^{W}$     & $R_C^{W}$     & $R_N^{W}$     & $R_O^{W}$    & 
  $R_F^{W}$     & $R_{Na}^{W}$  & $R_P^{W}$     & $R_S^{W}$    & 
  $R_{Cl}^{W}$  & $R_K^{W}$     & $R_{Br}^{W}$  & $R_J^{W}$    \\
  1.20          & 1.70          & 1.60          & 1.50         & 
  1.45          & 1.20          & 1.85          & 1.80         & 
  1.76          & 1.46          & 1.85          & 1.96         \\
\\ \\ \hline\hline
\end{tabular}
\end{center}
\end{table}
\clearpage
\pagebreak

\begin{table}[!htb]
\begin{center}
\thispagestyle{empty}
\caption[]{\label{table2} PDB codes and structural key data of a series 
                          of proteins used for comparison                 }
\vspace{0.5cm}
\begin{tabular}[t]{ccccc}
\\ \hline\hline \\
  Shape Sketch  & PDB-Code  & Number of    & Number of  & Charge  \\
                &           & Residues     & Atoms      & [a.u.]  \\
\\ \hline \\
  \raisebox{-0.3cm}{\includegraphics[scale=0.08]{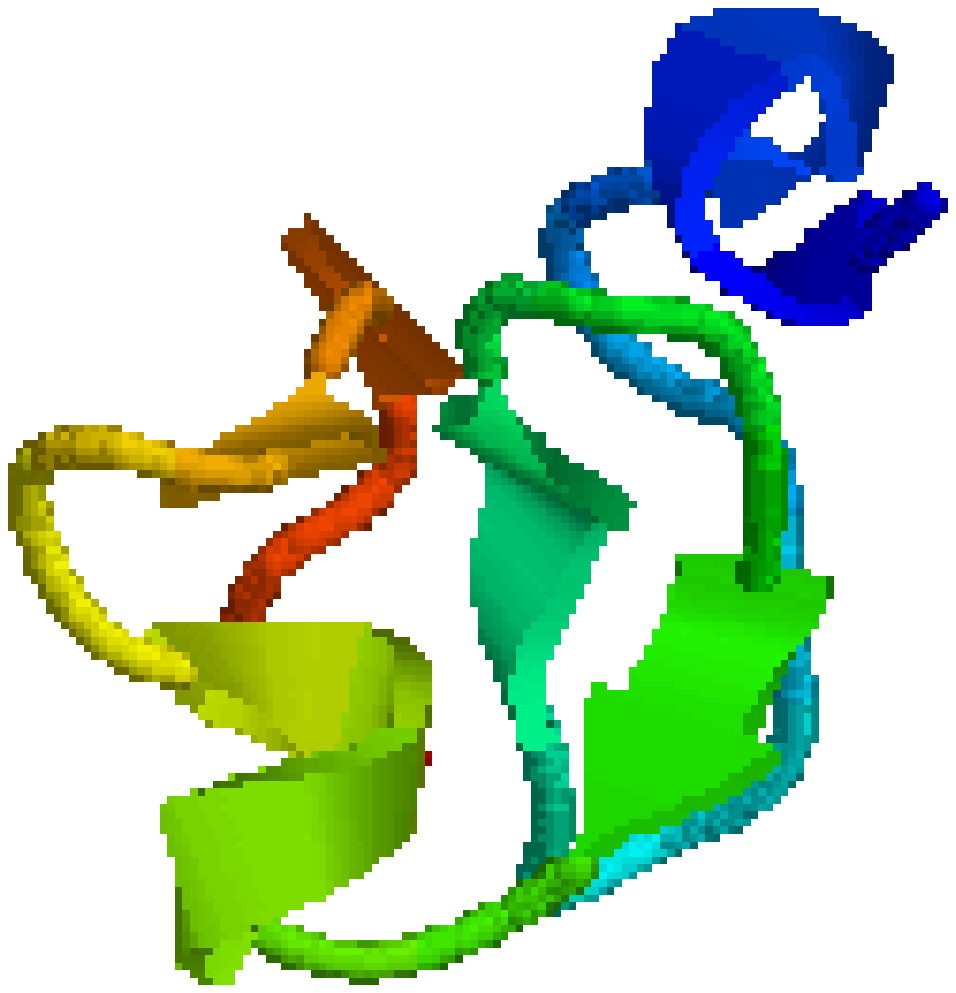}}  &
                  1P9GA     & 41           & 517        & +3      \\
  \raisebox{-0.3cm}{\includegraphics[scale=0.08]{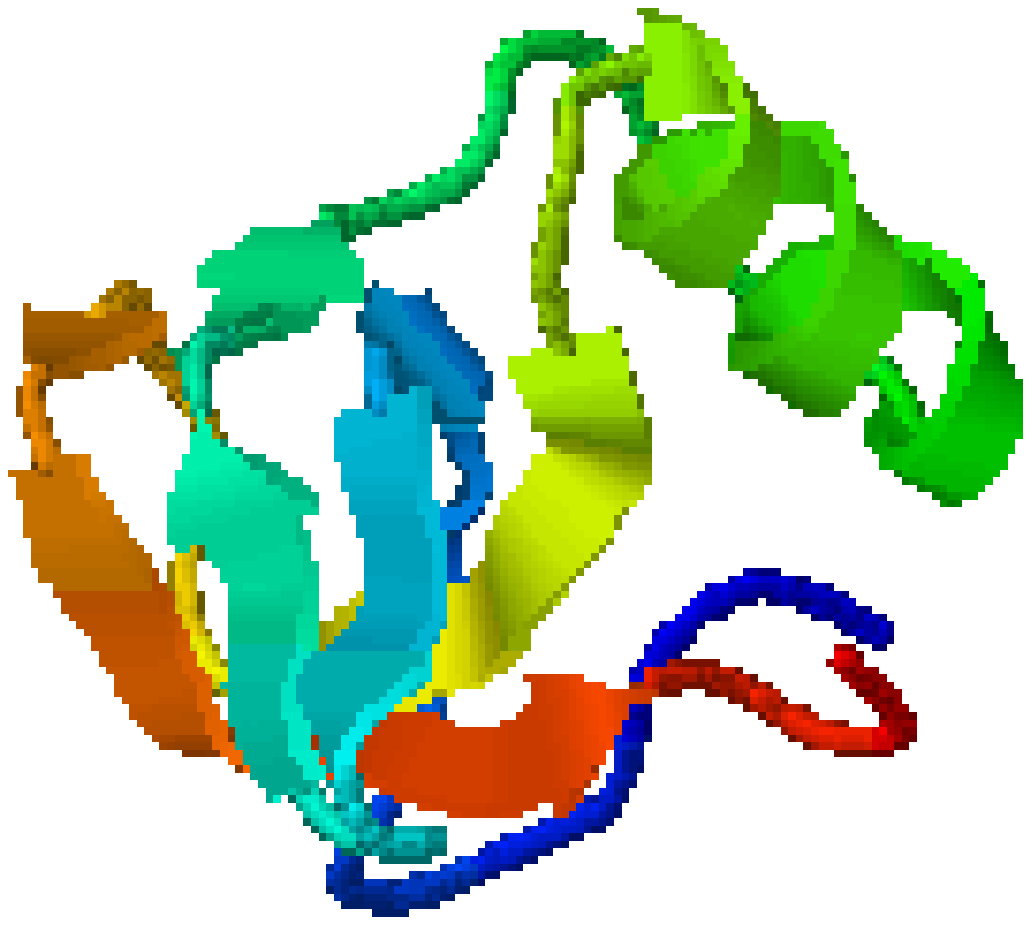}}   &
                  2B97      & 70           & 981        & +1      \\
  \raisebox{-0.3cm}{\includegraphics[scale=0.08]{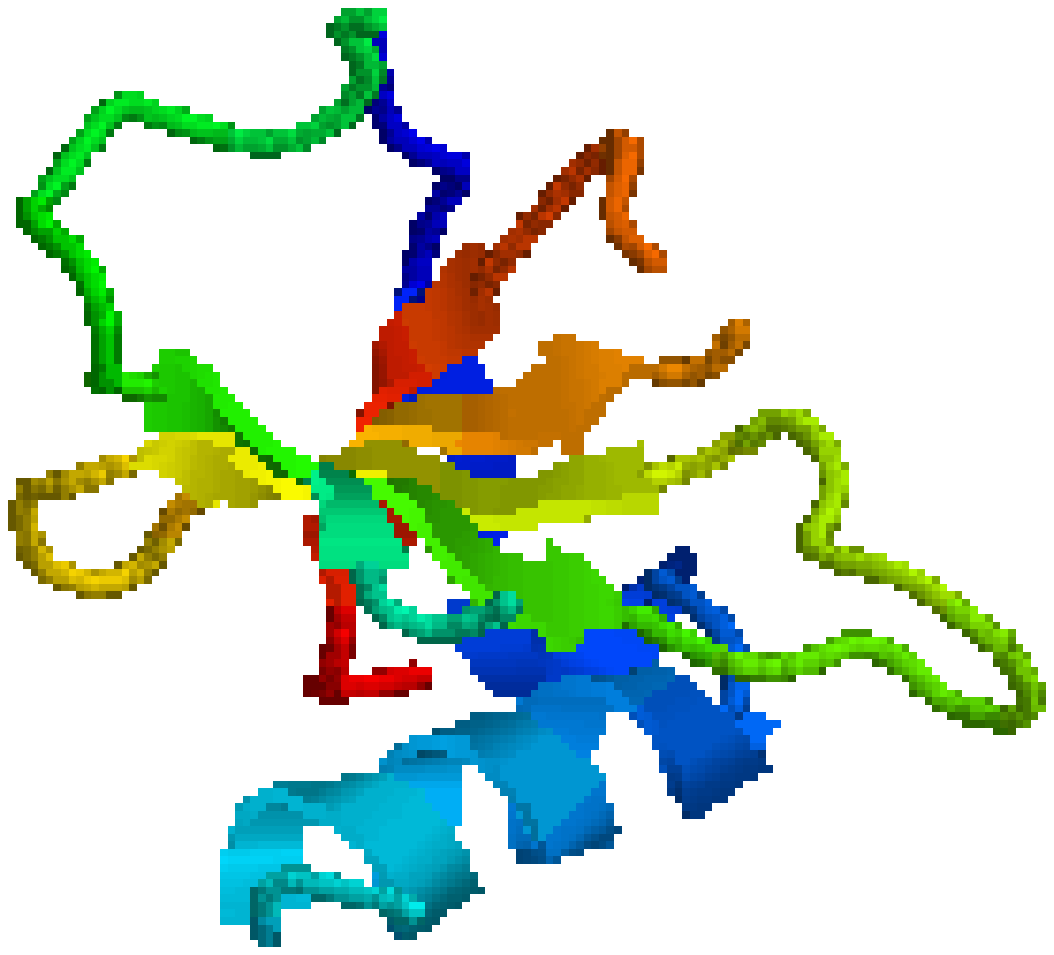}}   &
                  1LNI      & 96           & 1443       & -5      \\
  \raisebox{-0.3cm}{\includegraphics[scale=0.08]{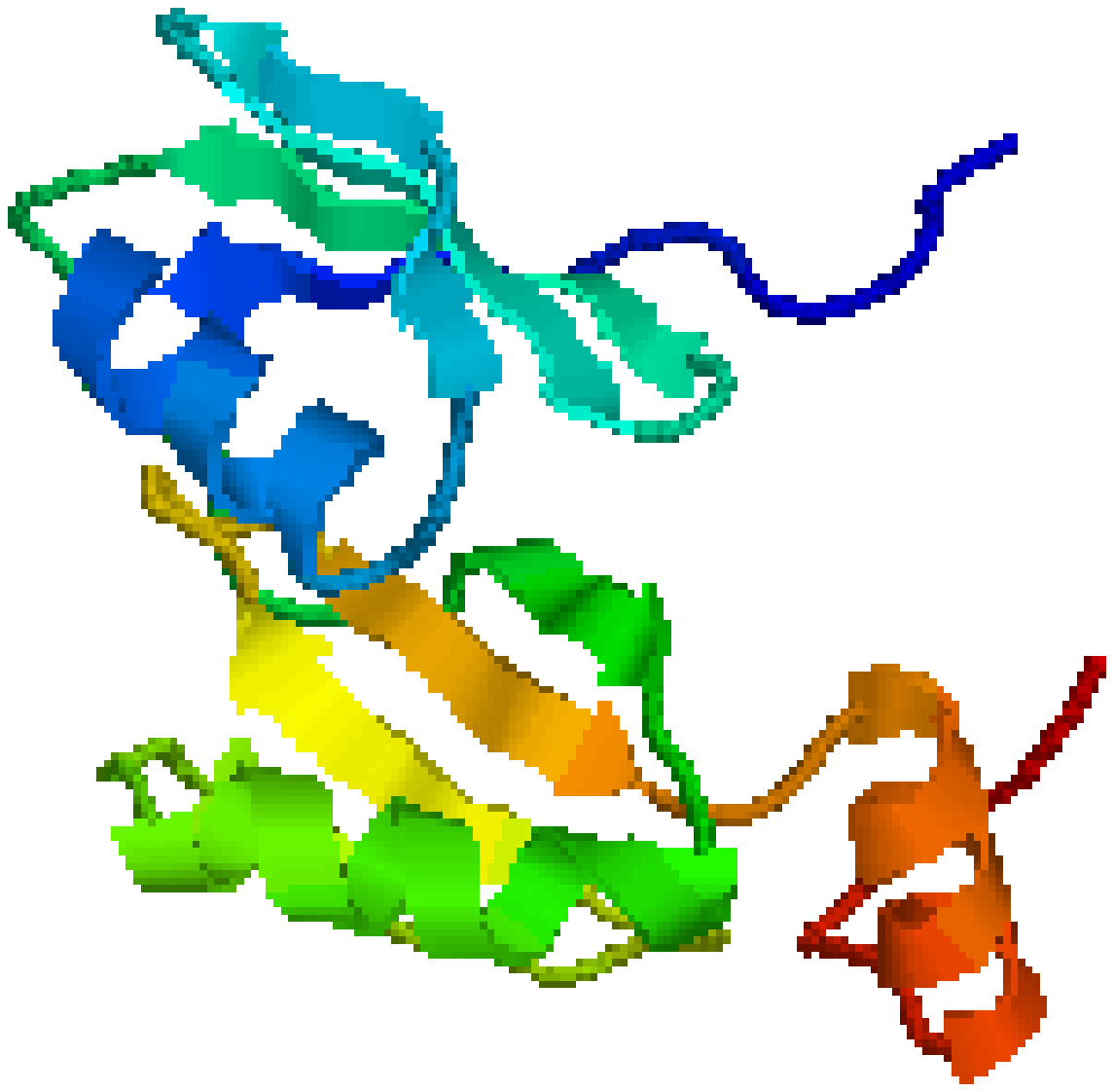}}   & 
                  1NKI      & 134          & 2082       & +5      \\
  \raisebox{-0.3cm}{\includegraphics[scale=0.08]{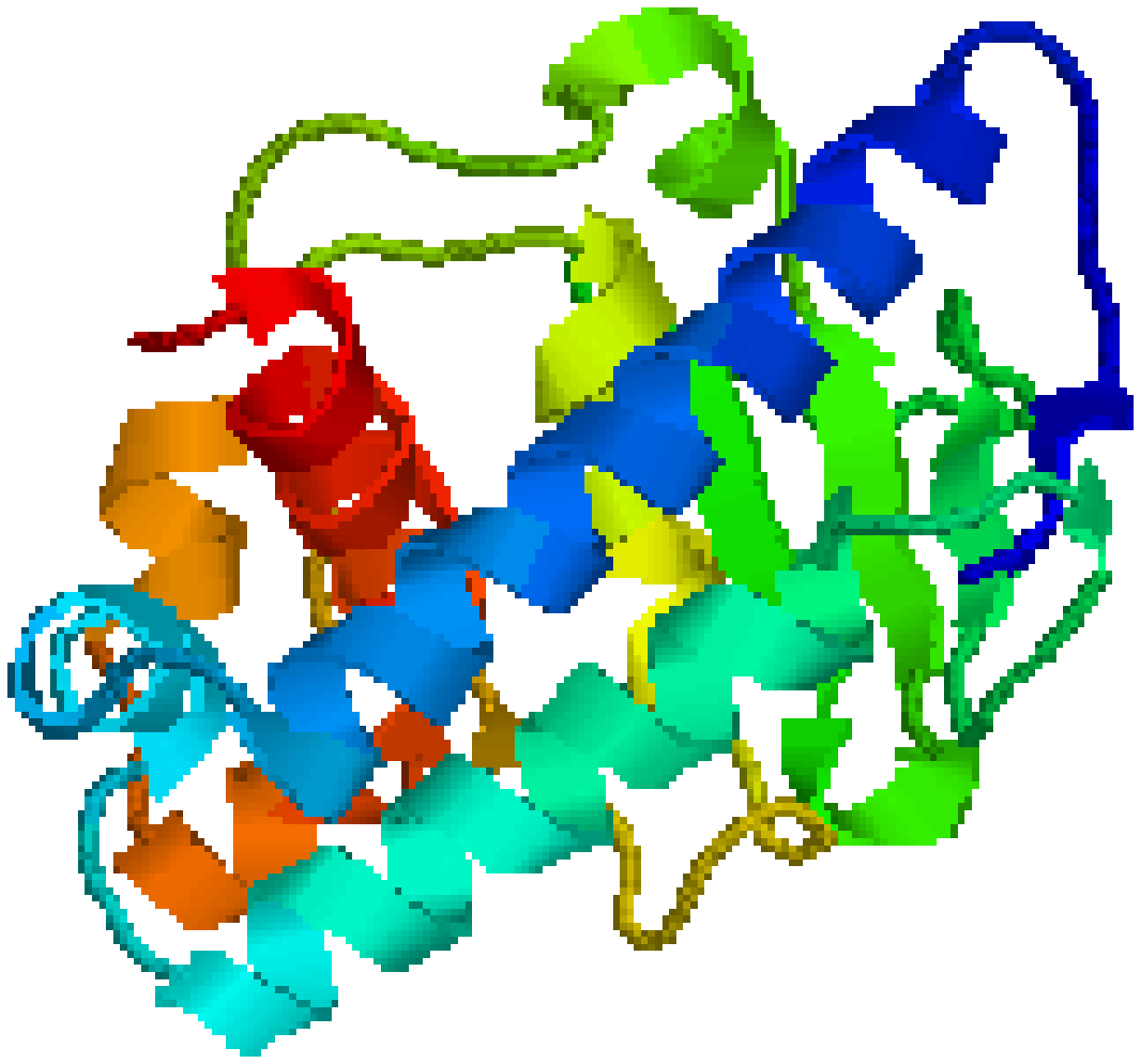}}   &
                  1EB6      & 177          & 2570       & -11     \\
  \raisebox{-0.3cm}{\includegraphics[scale=0.08]{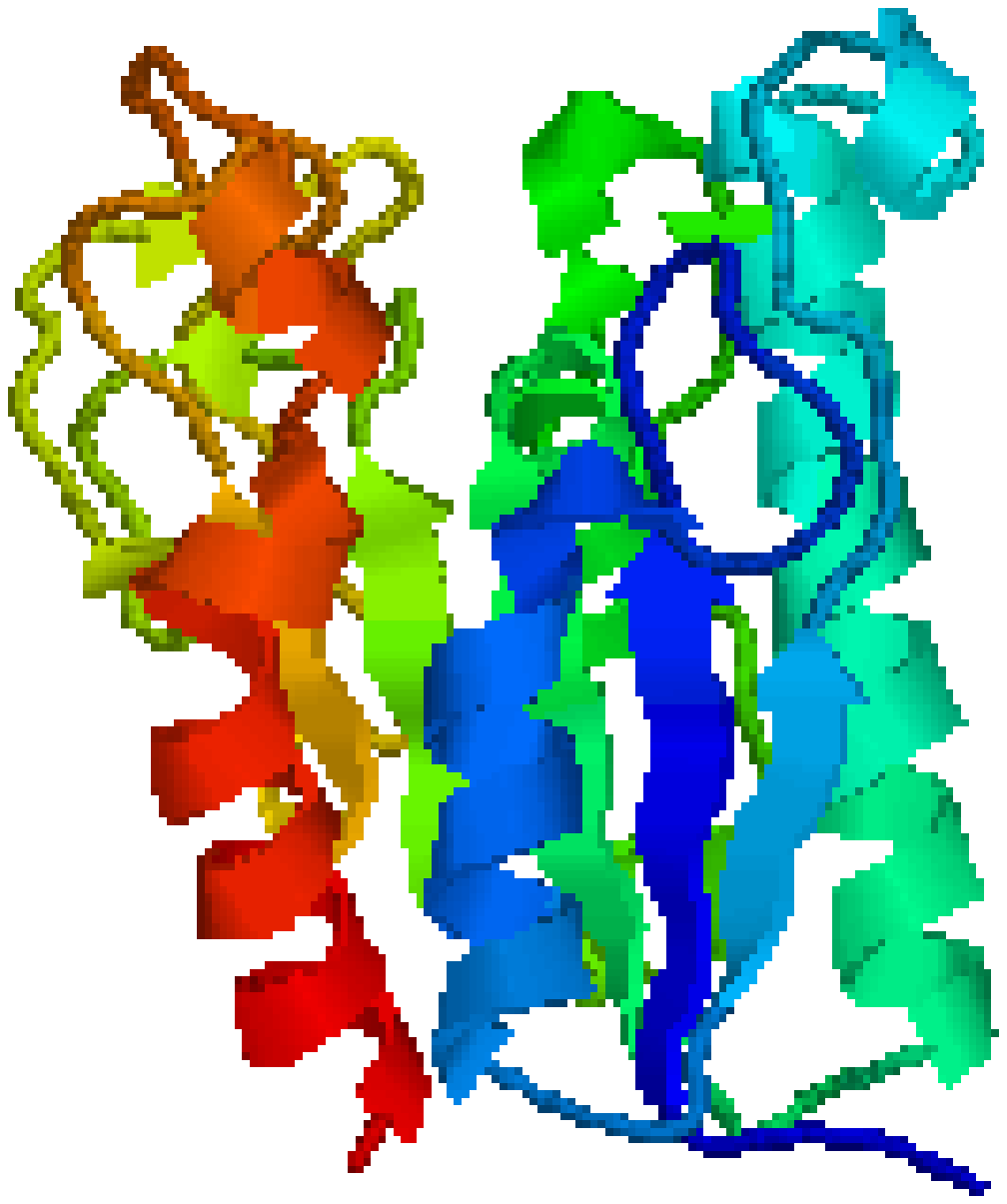}}   &
                  1G66      & 207          & 2777       & -2      \\
  \raisebox{-0.3cm}{\includegraphics[scale=0.08]{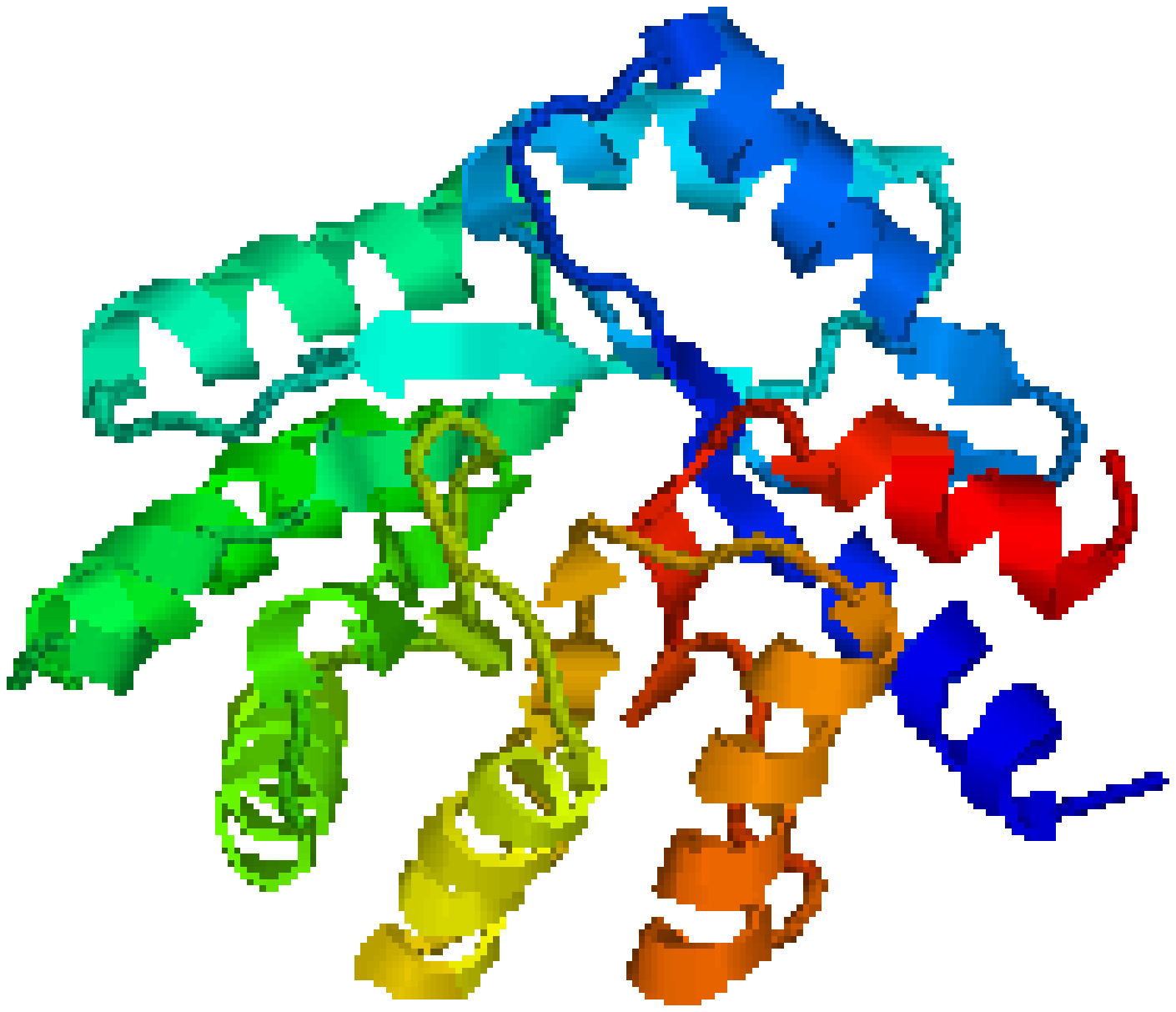}}   &
                  1P1X      & 250          & 3813       & 0       \\
  \raisebox{-0.3cm}{\includegraphics[scale=0.08]{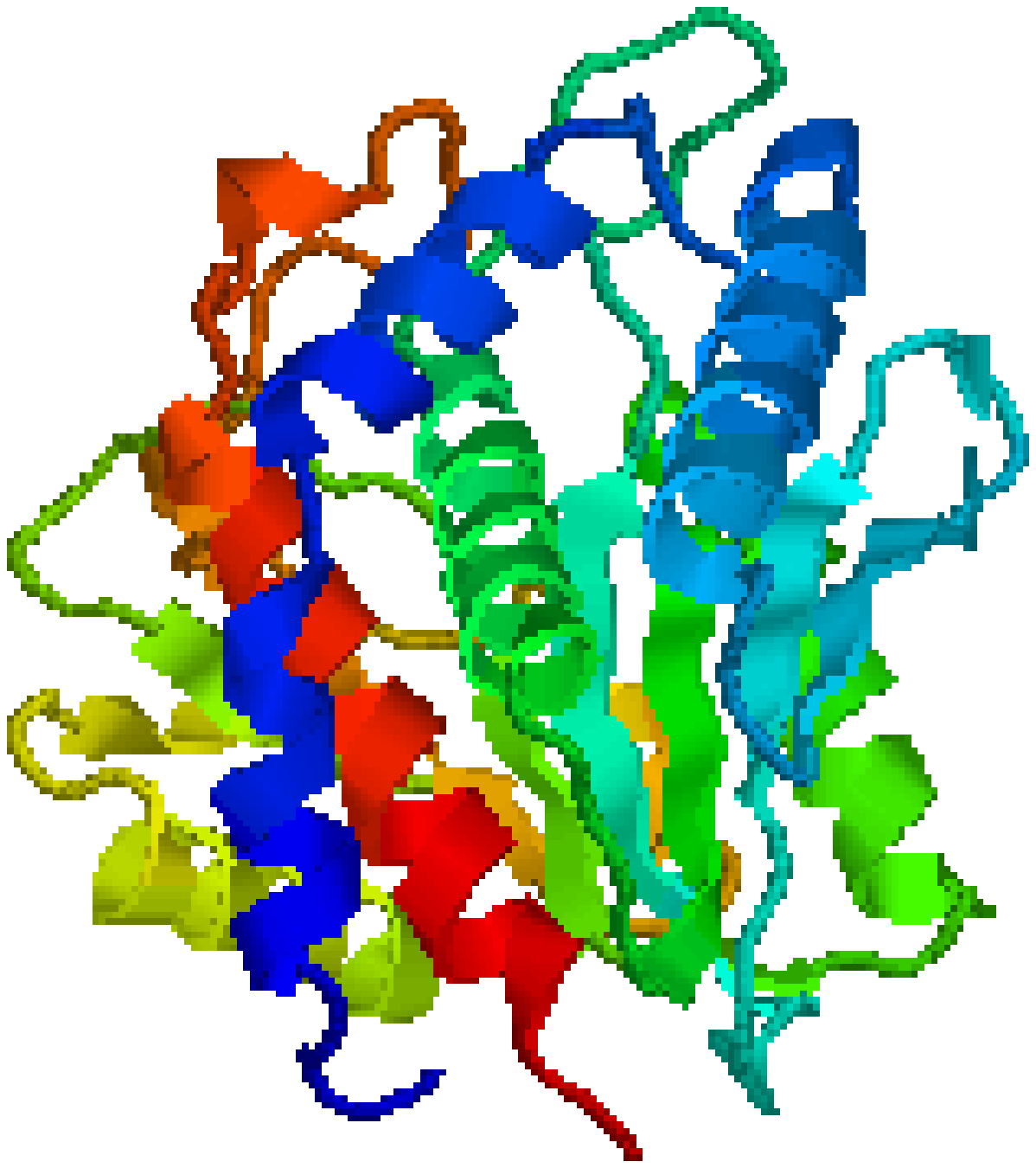}}   &
                  1RTQ      & 291          & 4287       & -16     \\
  \raisebox{-0.3cm}{\includegraphics[scale=0.08]{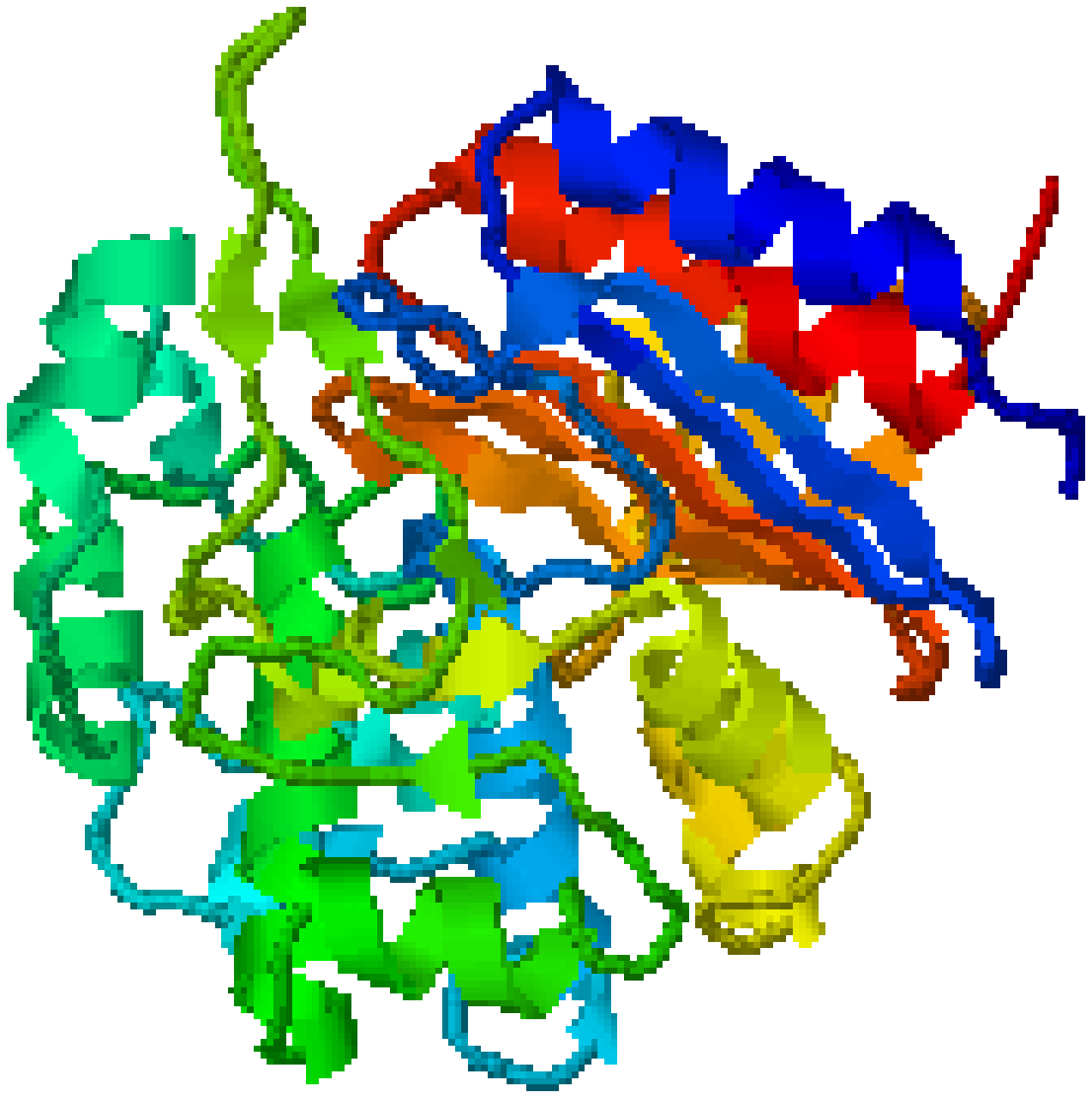}}   & 
                  1YQS      & 345          & 5147       & +2      \\
  \raisebox{-0.3cm}{\includegraphics[scale=0.08]{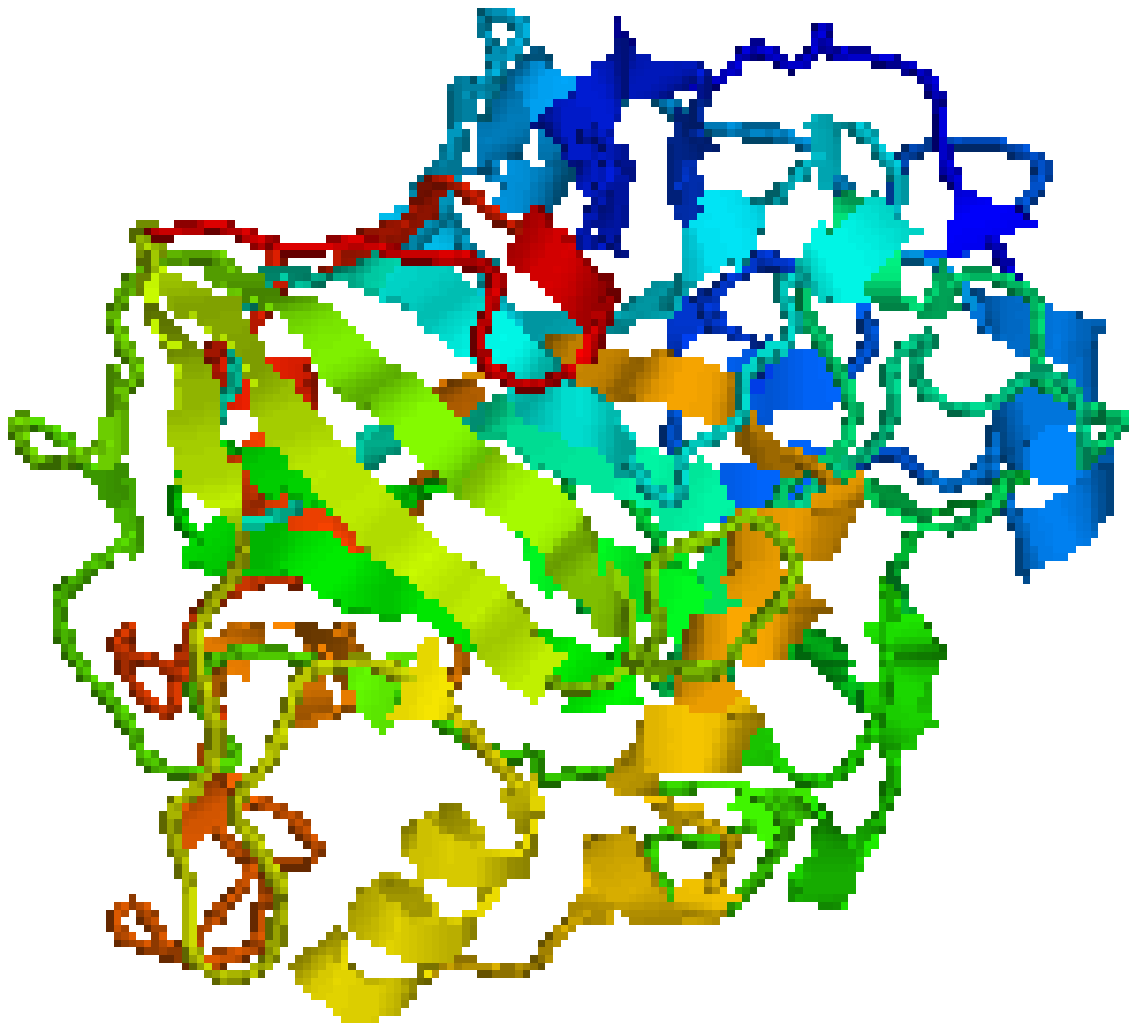}}   &
                  1GPI      & 430          & 6164       & -12     \\
\\ \\ \hline\hline
\end{tabular}
\end{center}
\end{table}
\clearpage
\pagebreak

\begin{table}[!htb]
\begin{center}
\caption[]{\label{table3} Comparison of PB/BEM-computed versus experimental 
                          total solvation free energies, $\Delta G^{solv}$, 
                          of amino acid side-chain analogues in water.
                          A scaling factor, $\lambda$, of 0.70 has been 
                          uniformly applied to all dispersion coefficients,
                          $\kappa_i$, in eq. \ref{eq2}. }
\vspace{0.5cm}
\begin{tabular}{lccc}
\\ \hline\hline \\
  Species                           &
  $\Delta G^{solv,PB/BEM}$          &  
  $\Delta G^{solv,Exp}$             & 
  Deviation                         \\
                                    &  
  $\left[\frac{kcal}{mol}\right]$   &  
  $\left[\frac{kcal}{mol}\right]$   &
  $\left[\frac{kcal}{mol}\right]$   \\
\\ \hline \\
  acetamide            &-10.97  & -9.68  & 1.29    \\
  butane               &  1.92  &  2.15  & 0.23    \\
  ethanol              & -4.58  & -4.88  & 0.30    \\
  isobutane            &  1.74  &  2.28  & 0.54    \\
  methane              &  0.72  &  1.94  & 1.22    \\
  methanethiol         & -3.57  & -1.24  & 2.33    \\
  methanol             & -6.58  & -5.06  & 1.52    \\
  methyl-ethyl-sulfide & -0.30  & -1.48  & 1.18    \\
  methylindole         & -4.19  & -5.88  & 1.69    \\
  p-cresol             & -3.56  & -6.11  & 2.55    \\
  propane              &  1.72  &  1.99  & 0.27    \\
  propionamide         & -9.34  & -9.38  & 0.04    \\
  toluene              &  1.05  & -0.76  & 1.81    \\
\\ \\ \hline\hline
\end{tabular}
\end{center}
\end{table}
\clearpage
\pagebreak

\begin{table}[!htb]
\begin{center}
\caption[]{\label{table4} Analysis of individual contributions to the 
                          net solvation free energy for solvent water
                          as computed by PB/BEM or by PCM and comparison 
                          to the experimental value.                      }
\vspace{0.5cm}
\begin{tabular}{lccccccccc}
\\ \hline\hline \\
  Species                           &
  $\Delta G^{cav}$                  &
  $\Delta G^{cav}$                  &
  $\Delta G^{disp}$                 &
  $\Delta G^{disp}_{rep}$           &
  $\Delta G^{pol}$                  &
  $\Delta G^{pol}$                  &
  $\Delta G^{solv}$                 &
  $\Delta G^{solv}$                 &
  $\Delta G^{solv}$                 \\
                                    &
  {\scriptsize PB/BEM }             &
  {\scriptsize PCM    }             &
  {\scriptsize PB/BEM }             &
  {\scriptsize PCM    }             &
  {\scriptsize PB/BEM }             &
  {\scriptsize PCM    }             &
  {\scriptsize PB/BEM }             &
  {\scriptsize PCM    }             &
  {\scriptsize Exp    }             \\
                                    &
  $\left[\frac{kcal}{mol}\right]$   &  
  $\left[\frac{kcal}{mol}\right]$   &
  $\left[\frac{kcal}{mol}\right]$   &
  $\left[\frac{kcal}{mol}\right]$   &
  $\left[\frac{kcal}{mol}\right]$   &
  $\left[\frac{kcal}{mol}\right]$   &
  $\left[\frac{kcal}{mol}\right]$   &
  $\left[\frac{kcal}{mol}\right]$   &
  $\left[\frac{kcal}{mol}\right]$   \\
\\ \hline \\
  acetamide            &   5.10  &  12.71 &  -4.26  & -7.45  &  
                         -11.81  & -14.13 & -10.97  & -8.88  & -9.68 \\
  butane               &   7.05  &  15.46 &  -4.41  & -8.48  &
                          -0.72  &  -0.45 &   1.92  &  6.54  &  2.15 \\
  ethanol              &   4.89  &  11.79 &  -3.71  & -6.74  &
                          -5.76  &  -6.42 &  -4.58  & -1.37  & -4.88 \\
  isobutane            &   7.17  &  15.94 &  -4.44  & -8.12  &
                          -0.98  &  -0.55 &   1.74  &  7.28  &  2.28 \\
  methane              &   3.08  &   9.98 &  -2.10  & -3.03  &
                          -0.26  &  -0.07 &   0.72  &  6.88  &  1.94 \\
  methanethiol         &   4.19  &  10.95 &  -4.12  & -6.77  &
                          -3.64  &  -4.35 &  -3.57  & -0.17  & -1.24 \\
  methanol             &   3.39  &   9.53 &  -3.05  & -4.88  &
                          -6.91  &  -6.02 &  -6.58  & -1.37  & -5.06 \\
  methyl-              &         &        &         &        &
                                 &        &         &        &       \\
  ethyl-sulfide        &   7.00  &  16.37 &  -5.30  & -9.49  &
                          -2.00  &  -3.02 &  -0.30  &  3.86  & -1.48 \\
  toluene              &   8.54  &  17.40 &  -5.51  &-11.17  &
                          -1.98  &  -3.73 &   1.05  &  2.51  & -0.76 \\
  methylindole         &  10.00  &  20.67 &  -7.09  &-14.10  &
                          -7.10  & -10.07 &  -4.19  & -3.50  & -5.88 \\
  p-cresol             &   8.92  &  18.93 &  -6.11  &-12.16  &
                          -6.37  & -10.48 &  -3.56  & -3.70  & -6.11 \\
  propane              &   5.80  &  13.58 &  -3.68  & -6.92  &
                          -0.40  &  -0.34 &   1.72  &  6.31  &  1.99 \\
  propionamide         &   6.34  &  14.56 &  -4.83  & -9.05  &
                         -10.84  & -13.05 &  -9.34  & -7.54  & -9.38 \\
\\ \\ \hline\hline
\end{tabular}
\end{center}
\end{table}
\clearpage
\pagebreak

\begin{table}[!htb]
\vspace*{-2cm}
\begin{center}
\caption[]{\label{table5} Individual contributions to the water net 
                          solvation free energy as computed from PB/BEM or 
                          PCM for a series of arbitrary small molecules
                          and comparison to the experimental value.       }
\vspace{0.5cm}
\begin{tabular}{lccccccccc}
\\ \hline\hline \\
  Species                           &
  $\Delta G^{cav}$                  &
  $\Delta G^{cav}$                  &
  $\Delta G^{disp}$                 &
  $\Delta G^{disp}_{rep}$           &
  $\Delta G^{pol}$                  &
  $\Delta G^{pol}$                  &
  $\Delta G^{solv}$                 &
  $\Delta G^{solv}$                 &
  $\Delta G^{solv}$                 \\
                                    &
  {\scriptsize PB/BEM }             &
  {\scriptsize PCM    }             &
  {\scriptsize PB/BEM }             &
  {\scriptsize PCM    }             &
  {\scriptsize PB/BEM }             &
  {\scriptsize PCM    }             &
  {\scriptsize PB/BEM }             &
  {\scriptsize PCM    }             &
  {\scriptsize Exp    }             \\
                                    &
  $\left[\frac{kcal}{mol}\right]$   &  
  $\left[\frac{kcal}{mol}\right]$   &
  $\left[\frac{kcal}{mol}\right]$   &
  $\left[\frac{kcal}{mol}\right]$   &
  $\left[\frac{kcal}{mol}\right]$   &
  $\left[\frac{kcal}{mol}\right]$   &
  $\left[\frac{kcal}{mol}\right]$   &
  $\left[\frac{kcal}{mol}\right]$   &
  $\left[\frac{kcal}{mol}\right]$   \\
\\ \hline \\
  propanal             &   6.04  &  13.26 &  -3.92  & -7.67  &  
                          -4.32  &  -6.35 &  -2.21  & -0.76  & -3.44 \\
  butanoic acid$^{(a)}$&   7.54  &  16.98 &  -5.31  &-10.3   &
                          -8.18  & -10.85 &  -5.94  & -4.16  & -6.47 \\
  cyclohexane          &   8.77  &  16.45 &  -5.29  &-11.56  &
                           0.00  &  -0.58 &   3.48  &  4.31  &  1.23 \\
  acetone              &   5.77  &  14.30 &  -3.96  & -7.04  &
                          -4.67  &  -6.05 &  -2.85  &  1.21  & -3.85 \\
  propene              &   5.55  &  12.60 &  -3.58  & -6.48  &     
                          -0.98  &  -1.24 &   0.99  &  4.88  &  1.27 \\
  propionic acid$^{(a)}$&  6.31  &  14.57 &  -4.60  & -8.73  &
                          -8.38  & -10.42 &  -6.67  & -4.59  & -6.47 \\
  propyne              &   4.87  &  12.07 &  -3.14  & -5.75  &
                          -2.36  &  -3.33 &  -0.62  &  2.99  & -0.31 \\
  hexanoic acid$^{(a)}$&   9.97  &        &  -6.76  &        &
                          -8.33  &        &  -5.12  &        & -6.21 \\
  anisole              &   8.66  &        &  -6.12  &        &
                          -3.27  &        &  -0.73  &        & -2.45 \\
  benzaldehyde         &   8.47  &  17.26 &  -5.74  &-11.99  &
                          -5.05  &  -9.38 &  -2.32  & -4.12  & -4.02 \\
  ethyne               &   4.16  &   9.78 &  -2.76  & -4.92  &
                          -0.96  &  -1.05 &   0.44  &  3.81  &  1.27 \\
  butanal              &   7.18  &  15.75 &  -4.63  & -9.33  &
                          -4.55  &  -6.77 &  -2.00  & -0.36  & -3.18 \\
  benzene              &   7.24  &  14.21 &  -4.84  &-10.27  &
                          -2.76  &  -4.04 &  -0.36  & -0.10  & -0.87 \\
  bromobenzene         &   8.67  &  16.96 &  -5.91  &-12.73  &
                          -2.46  &  -4.76 &   0.29  & -0.53  & -1.46 \\
  acetic acid$^{(a)}$  &   4.89  &  12.38 &  -3.92  & -7.02  &
                          -8.41  & -10.49 &  -7.44  & -5.13  & -6.70 \\
  bromoethane          &   5.95  &  13.09 &  -4.25  & -8.35  &
                          -1.61  &  -2.77 &   0.09  &  1.98  & -0.70 \\
  ethylbenzene         &   9.57  &  19.57 &  -6.11  &-12.68  &
                          -1.92  &  -3.62 &   1.54  &  3.27  & -0.80 \\
  diethylether         &   7.49  &  17.71 &  -5.12  & -9.47  &
                          -1.41  &  -2.48 &   0.97  &  5.76  & -1.76 \\
\\ \\ \hline\hline
\end{tabular}
\end{center}
{\scriptsize $^{(a)}$ protonated form }
\end{table}
\clearpage
\pagebreak

\begin{table}[!htb]
\begin{center}
\caption[]{\label{table6} Comparison of PB/BEM computed solvation 
                          free energies of zwitterionic amino acids in 
                          water against data by Chang et al. \cite{chang}
                          obtained from Monte Carlo Free Energy 
                          simulations. Dispersion coefficients are 
                          scaled by the multiplicative factor of 0.70    
                          in PB/BEM.                                       }
\vspace{0.5cm}
\begin{tabular}{lccccccccc}
\\ \hline\hline \\
  Species                           &
  $\Delta G^{solv,PB/BEM}$          &  
  $\Delta G^{solv,MC}$              & 
  Deviation                         \\
                                    &  
  $\left[\frac{kcal}{mol}\right]$   &
  $\left[\frac{kcal}{mol}\right]$   &
  $\left[\frac{kcal}{mol}\right]$   \\
\\ \hline \\
  Gly                  & -55.73  & -56.80  &  1.07     \\
  Ala                  & -51.75  & -57.70  &  5.95     \\
  Val                  & -48.79  & -56.20  &  7.41     \\
  Leu                  & -49.05  & -57.30  &  8.25     \\
  Ile                  & -47.55  & -55.70  &  8.15     \\
  Ser                  & -60.82  & -55.30  &  5.52     \\
  Thr                  & -61.33  & -54.40  &  6.93     \\
  Cys                  & -60.86  & -54.70  &  6.16     \\
  Met                  & -50.88  & -57.30  &  6.42     \\
  Asn                  & -58.63  & -60.10  &  1.47     \\
  Gln                  & -65.82  & -59.60  &  6.22     \\
  Phe                  & -51.46  & -55.90  &  4.44     \\
  Tyr                  & -55.16  & -61.60  &  6.44     \\
  Trp                  & -58.00  & -64.60  &  6.60     \\
\\ \\ \hline\hline
\end{tabular}
\end{center}
\end{table}
\clearpage
\pagebreak

\begin{table}[!htb]
\begin{center}
\caption[]{\label{table7} Comparison of PB/BEM-computed versus experimental 
                          total solvation free energies, $\Delta G^{solv}$, 
                          of various substances in ethanol. A scaling factor, 
                          $\lambda$, of 0.82 has been uniformly applied to 
                          all dispersion coefficients, $\kappa_i$, in eq. 
                          \ref{eq2}.                                         }
\vspace{0.5cm}
\begin{tabular}{lccc}
\\ \hline\hline \\
  Species                           &
  $\Delta G^{solv,PB/BEM}$          &  
  $\Delta G^{solv,Exp}$             & 
  Deviation                         \\
                                    &  
  $\left[\frac{kcal}{mol}\right]$   &  
  $\left[\frac{kcal}{mol}\right]$   &
  $\left[\frac{kcal}{mol}\right]$   \\
\\ \hline \\
  n-octane             & -0.70  & -4.23  & 3.53    \\
  toluene              & -3.30  & -4.57  & 1.27    \\
  dioxane              & -6.03  & -4.68  & 1.35    \\
  butanone             & -4.83  & -4.32  & 0.51    \\
  chlorobenzene        & -3.52  & -3.30  & 0.22    \\
\\ \\ \hline\hline
\end{tabular}
\end{center}
\end{table}
\clearpage
\pagebreak

\begin{table}[!htb]
\begin{center}
\caption[]{\label{table8} Comparison of PB/BEM-computed versus experimental 
                          total solvation free energies, $\Delta G^{solv}$, 
                          of various substances in n-octanol. A scaling 
                          factor, $\lambda$, of 0.74 has been uniformly 
                          applied to all dispersion coefficients, 
                          $\kappa_i$, in eq.  \ref{eq2}.                    }
\vspace{0.5cm}
\begin{tabular}{lccc}
\\ \hline\hline \\
  Species                           &
  $\Delta G^{solv,PB/BEM}$          &  
  $\Delta G^{solv,Exp}$             & 
  Deviation                         \\
                                    &  
  $\left[\frac{kcal}{mol}\right]$   &  
  $\left[\frac{kcal}{mol}\right]$   &
  $\left[\frac{kcal}{mol}\right]$   \\
\\ \hline \\
  acetone                & -5.28  & -3.15  & 2.13    \\
  anisole                & -4.80  & -5.47  & 0.67    \\
  benzaldehyde           & -6.16  & -6.13  & 0.03    \\
  benzene                & -3.87  & -3.72  & 0.15    \\
  bromobenzene           & -3.75  & -7.47  & 3.72    \\
  butanal                & -5.02  & -4.62  & 0.40    \\
  butanoic acid$^{(a)}$  & -8.74  & -7.58  & 1.16    \\
  cyclohexane            & -0.64  & -3.46  & 2.82    \\
  acetic acid$^{(a)}$    & -8.96  & -6.35  & 2.61    \\
  ethylbenzene           & -2.94  & -5.08  & 2.14    \\
  ethylene               & -1.57  & -0.27  & 1.30    \\
  hexanoic acid$^{(a)}$  & -8.89  & -8.82  & 0.07    \\
  propanal               & -4.71  & -4.13  & 0.58    \\
  propionic acid$^{(a)}$ & -8.75  & -6.86  & 1.89    \\
  propene                & -1.61  & -1.14  & 0.47    \\
  propyne                & -2.81  & -1.59  & 1.22    \\
  bromoethane            & -2.69  & -2.90  & 0.21    \\
\\ \\ \hline\hline
\end{tabular}
\end{center}
{\scriptsize $^{(a)}$ protonated form }
\end{table}
\clearpage
\pagebreak

\begin{table}[!htb]
\begin{center}
\caption[]{\label{table9} Summary of optimized parameters to be used
                          in PB/BEM for different types of solvents.
                          \textcolor{red}{ Average sizes of BEs are given 
                                           as pairs of values employed for 
                                           calculation of $\Delta G^{pol}$ 
                                           and $\Delta G^{disp}$ 
                                           respectively.
                                         }                                   }
\vspace{0.5cm}
\begin{tabular}{lcccc}
\\ \hline\hline \\
  Parameter Class                   &
  Water                             &  
  Methanol                          & 
  Ethanol                           & 
  n-Octanol                         \\
\\ \hline \\
  BE Average Size [\AA$^2$]         & 0.31\textcolor{red}{/0.45}  & 
                                      0.31\textcolor{red}{/0.45}  & 
                                      0.31\textcolor{red}{/0.45}  &  
                                      0.31\textcolor{red}{/0.45}  \\
  Probe Sphere Radius [\AA]         & 1.50   & 1.90   & 2.20 &  2.945  \\
  AMBER vdW Radii Scaling           & 1.07   & 1.06   & 1.06 &  1.05   \\
  AMBER Partial Charges Scaling     & 1.00   & 1.00   & 1.00 &  1.00   \\
  Caillet-Claverie Dispersion       &        &        &                \\
  Coefficients Scaling              & 0.70   & --     & 0.82 &  0.74   \\
  \textcolor{red}{ AMBER vdW Potential } &   &        &                \\
  \textcolor{red}{ Well Depth Scaling  } & 
  \textcolor{red}{ 0.76 }           & 
  \textcolor{red}{  --  }           & 
  \textcolor{red}{ 0.94 }           &  
  \textcolor{red}{ 2.60 }           \\
\\ \\ \hline\hline
\end{tabular}
\end{center}
\end{table}
\clearpage
\pagebreak

\begin{table}[!htb]
\vspace*{-2cm}
\textcolor{red}{
                \begin{center}
                \caption[]{\label{table10} Effect on total solvation free 
                                           energies for water as 
                                           PB/BEM-computed with AMBER style of 
                                           dispersion (eq. \ref{eq3}) versus 
                                           Caillet-Claverie style of dispersion
                                           (eq. \ref{eq2}) and comparison to
                                           the experimental value.       } 
                \vspace{0.5cm}
                \begin{tabular}{lccc}
                \\ \hline\hline \\
                  Species                               &
                  $\Delta G^{solv}_{Caillet-Claverie}$  &
                  $\Delta G^{solv}_{AMBER}$             &
                  $\Delta G^{solv}_{Exp}$               \\
                                                    &  
                  $\left[\frac{kcal}{mol}\right]$   &
                  $\left[\frac{kcal}{mol}\right]$   &
                  $\left[\frac{kcal}{mol}\right]$   \\
                \\ \hline \\
                  propanal              &   -2.21 &  -1.71 &  -3.44  \\            
                  butanoic acid$^{(a)}$ &   -5.94 &  -6.00 &  -6.47  \\           
                  cyclohexane           &    3.48 &  -1.33 &   1.23  \\          
                  acetone               &   -2.85 &  -2.42 &  -3.85  \\          
                  propionic acid$^{(a)}$&   -6.67 &  -6.57 &  -6.47  \\             
                  propyne               &   -0.62 &  -2.09 &  -0.31  \\           
                  hexanoic acid$^{(a)}$ &   -5.12 &  -5.81 &  -6.21  \\        
                  anisole               &   -0.73 &  -3.49 &  -2.45  \\        
                  benzaldehyde          &   -2.32 &  -3.22 &  -4.02  \\           
                  butanal               &   -2.00 &  -1.86 &  -3.18  \\          
                  benzene               &   -0.36 &  -2.78 &  -0.87  \\          
                  bromobenzene          &    0.29 &  -1.63 &  -1.46  \\          
                  acetic acid$^{(a)}$   &   -7.44 &  -6.76 &  -6.70  \\              
                  bromoethane           &    0.09 &  -0.32 &  -0.70  \\            
                  ethylbenzene          &    1.54 &  -1.25 &  -0.80  \\          
                  diethylether          &    0.97 &  -0.68 &  -1.76  \\          
                \\ \\ \hline\hline
                \end{tabular}
                \end{center}
                {\scriptsize $^{(a)}$ protonated form }
               }
\end{table}
\clearpage
\pagebreak

\begin{table}[!htb]
\vspace*{-2cm}
\begin{center}
\caption[]{\label{table11} Analysis of partial term contributions to  
                           PB/BEM-computed solvation free energies for
                           a series of proteins of increasing size using
                           either AMBER standard partial charges or 
                           semi-empirical PM5 charges obtained from 
                           program LocalSCF \cite{anikin}. COSMO 
                           approximations obtained from LocalSCF are
                           included.                                       }
\vspace{0.5cm}
\begin{tabular}{lcccccccc}
\\ \hline\hline \\
  Species                               &
  $\frac{Surface}{Volume}$              &
  $\Delta G^{cav}$                      &
  $\Delta G^{disp}$                     &
  $\Delta G^{pol}_{AMBER}$              &
  $\Delta G^{solv}_{AMBER}$             &
  $\Delta G^{pol}_{PM5}$                &
  $\Delta G^{solv}_{PM5}$               &
  $\Delta G^{solv}_{COSMO,PM5}$         \\
                                    &  
  $\left[\frac{1}{\AA}\right]$      &
  $\left[\frac{kcal}{mol}\right]$   &
  $\left[\frac{kcal}{mol}\right]$   &
  $\left[\frac{kcal}{mol}\right]$   &
  $\left[\frac{kcal}{mol}\right]$   &
  $\left[\frac{kcal}{mol}\right]$   &
  $\left[\frac{kcal}{mol}\right]$   &
  $\left[\frac{kcal}{mol}\right]$   \\
\\ \hline \\
  1P9GA & 0.42  & 114.6   &  -66.2     &  -339.9 &  -291.6 &  
         -251.6 &  -203.3 &    36849.9 \\
  2B97  & 0.34  & 181.3   &  -91.3     &  -548.0 &  -457.9 &  
         -517.5 &  -427.4 &    55374.0 \\
  1LNI  & 0.34  & 237.5   & -126.8     & -1418.6 & -1307.9 & 
        -1140.3 & -1029.6 &   262801.0 \\
  1NKI  & 0.37  & 305.4   & -199.2     & -1652.1 & -1546.0 & 
        --      & --      &   224844.2 \\
  1EB6  & 0.28  & 353.0   & -182.7     & -2571.9 & -2401.7 & 
        -2312.2 & -2141.9 &  -768986.4 \\
  1G66  & 0.26  & 369.0   & -187.5     & -1193.6 & -1012.1 &  
         -881.5 &  -700.0 &  -259150.6 \\
  1P1X  & 0.25  & 459.3   & -235.4     & -2434.6 & -2210.7 & 
        -2106.8 & -1882.9 &    17998.3 \\
  1RTQ  & 0.22  & 506.4   & -238.2     & -4077.5 & -3809.3 & 
        -3172.4 & -2904.1 & -1604637.0 \\
  1YQS  & 0.23  & 566.9   & -286.4     & -2133.4 & -1852.9 & 
        -1680.8 & -1400.3 &   332117.0 \\
  1GPI  & 0.24  & 651.8   & -342.6     & -3961.3 & -3652.1 & 
        -3252.0 & -2942.8 & -1605637.8 \\
\\ \\ \hline\hline
\end{tabular}
\end{center}
\end{table}
\clearpage
\pagebreak

\listoffigures
\clearpage
\pagebreak

\clearpage
\pagebreak

\clearpage
\pagebreak

\clearpage
\pagebreak

\clearpage
\pagebreak

\clearpage
\pagebreak

\end{document}